\magnification=1200
\parskip=10pt plus 5pt
\parindent=15pt
\baselineskip=15pt
\pageno=0
\footline={\ifnum \pageno <1 \else \hss \folio \hss \fi}
\topglue 1.3in
\centerline{\bf Effectively Closed Infinite-Genus Surfaces and the String 
Coupling}
\vskip .6in
\centerline{\bf Simon Davis}
\vskip .6in
\centerline{\bf Institut f{\"u}r Theoretische Physik}
\centerline{\bf Freie Universit{\"a}t Berlin}
\centerline{\bf Arnimallee 14}
\centerline{\bf 14195 Berlin}
\vskip .4in
\noindent
{\bf Abstract.} The class of effectively closed infinite-genus surfaces, 
defining the completion of the domain of string perturbation theory, can be 
included in the category $O_G$, which is characterized by the vanishing capacity
of the ideal boundary.  The cardinality of the maximal set of endpoints
is shown to be $2^{\mit N}$.  The product of the coefficient of the 
genus-$g$ superstring amplitude in four dimensions by $2^g$ in the 
$g\to \infty$ limit is an exponential function of the genus with a base
comparable in magnitude to the unified gauge coupling.  The value of
the string coupling is consistent with the characteristics of configurations
which provide a dominant contribution to a finite vacuum amplitude. 
\vfill
\eject
\noindent
{\bf 1. Introduction.}  
\vskip 5pt
The vanishing of multi-loop superstring vacuum amplitudes at finite order in
perturbation theory implies that there must be an additional contribution to
the inner product
$$\langle  0_{out}\vert S \vert 0_{in}\rangle~=~\langle 0_{in} 
\vert S^{\dag} S \vert 0_{in}\rangle~=~1
\eqno(1.1)
$$
given a unitary S-matrix.  The contribution of disconnected bubble diagrams, 
which is nevertheless factored from the vacuum amplitude, also would be 
zero, since the contour arguments in the non-renormalization theorems for
superstrings continue to be valid.

Additional contributions to the amplitude could arise from surfaces with
boundaries with non-zero linear measure and punctures.  However, these 
boundaries occur either when the theory contains both open and closed strings 
or through non-perturbative effects, whereas punctures are associated with 
additional asymptotic string states. 

It follows that the domain of closed string perturbation theory should be 
extended to infinite-genus super-Riemann surfaces to calculate the vacuum amplitude.
The geometrical conditions imposed on this space will be
defined first with respect to the ordinary part of the supermanifold, 
since there are no additional discrete modular
transformations in the odd Grassmann coordinates.  As the interaction 
region in a scattering process has finite volume, the surfaces should be 
effectively closed, and they share the property that they can be 
uniformized by Schottky groups, except that the number of generators 
may be infinite.  Specifically, effectively closed surfaces will have 
handles with thickness decreasing at a rate ${1\over {n^q}},~q>{1\over 2}$, 
so that they are conformally equivalent to spheres with handles accumulating 
to a point even in the intrinsic hyperbolic metric.  These 
surfaces would be required to satisfy the property of conformal rigidity, 
since it should not be possible to map them to manifolds with a boundary 
at infinity having non-zero linear measure, which could not contribute to the
perturbative expansion of the S-matrix element representing the amplitude
for the $n_1$ in-states and $n_2$ out-states for fixed $n_1,~n_2$ [1].

While the bosonic Polyakov action is known as the energy functional for the
harmonic map $z\to X^\mu(z)$ from a Riemann surface to the target space, its
extremal value is the area in the intrinsic metric and not the total energy
of the scattering process.  Infinite-genus surfaces can be included in
the perturbative expansion of the S-matrix because their infinite area in
the hyperbolic metric is associated with a vanishing weighting 
factor compensated by an integral over an infinite-dimensional moduli 
space. 

The structure of ${\mit M}_\infty$ within the category of effectively closed
infinite-genus surfaces is described. After determining the appropriate 
normalization of the superstring amplitudes, it is verified that the volume 
of the supermoduli space integral increases exponentially with the genus.
The functional dependence on the string coupling $\kappa_{str.}$ can be 
estimated once accurate bounds are placed on the volume integral.  In the 
Schottky parameterization, the primitive-element products and the determinant 
factor depend exponentially on the genus, whereas, in the Fuchsian 
parameterization, the integrand is expressed in terms of Selberg zeta 
functions, with the dominant behaviour given by the lowest eigenvalues of the 
Laplacian operator.

The ends of an infinite-genus surface can be identified generally with
the points in a Cantor set. Defining the cardinality of the set of ends 
to be $card~E$, finiteness of the infinite-genus contribution to the 
superstring amplitude implies the condition  
$$(card~E)lim_{g\to \infty} c(\kappa_{str})^g <\infty
\eqno(1.2)
$$
where $c(\kappa_{str})^g$ is the exponential genus-dependence of the 
supermoduli space integral.  This relation would determine a specific value 
for string coupling, which then can be compared to the gauge coupling
of supersymmetric grand unified theories.
\vskip 10pt
\noindent
{\bf 2. The Domain of the String Path Integral}
\vskip 5pt
\noindent
{\bf 2.1. Universal Moduli Space}
\vskip 5pt
The closure of moduli space is ${\bar {\mit M}}_g={\mit M}_g\cup {\mit D}_g$
where ${\mit D}_g$ is the compactification divisor at genus g consisting of
degenerate Riemann surfaces.  Universal moduli space can be defined to be
${\bar {\mit R}}=\prod_{g=0}^\infty~\left(\cup_{k=0}^\infty~Sym^k {\bar 
{\mit M}}_g\right)$ or ${\bar {\mit R}}_\infty={\mit R}\times
\cup_{k=0}^\infty~Sym^k({\bar {\mit M}}_\infty)$, where 
${\bar {\mit M}}_\infty$ is the space of connected, stable, essentially
compact surfaces of infinite genus [2], and the Sym symbol is 
included so that there is no overcounting of complex structures on two 
separate genus-$g$ components.

Almost all of the handles on infinite-genus surfaces in
${\bar {\mit M}_\infty}$ will be very small.  According to the geometric
hypotheses for the definition of Riemann surfaces [3], they can be 
constructed by applying glueing maps to copies of ${\bf C}$ with 
open, simply connected neighbourhoods $S_\nu$ around a set of points and 
an additional compact set $K_j$ deleted.  The glueing maps are constrained 
by the inequalities
$$\eqalign{
R_\mu(j) &< {1\over 4}~min_{{s\in S_{\nu_\mu}}\atop {s\neq 
s_\mu(j)}}~\vert s-s_\mu(j)\vert~~~~~~~~~~s_\mu(j)\in  S_\mu
\cr
R_\mu(j) &< {1\over 4}~dist (s_\mu(j), K_{\nu_\mu}(j))
\cr
\sum_{j,\mu}~{1\over {\vert s_\mu(j)\vert^{d-4\delta-2}}}~&<~\infty
\cr
r_\mu(j)&< {1\over {\vert s_\mu(j)\vert^d}}~~~~~~~~~R_\mu(j)
> {1\over {\vert s_\mu(j)\vert^\delta}}
\cr
\vert s_1(j)-s_2(j)\vert &> {1\over {\vert s_\mu(j)\vert^\delta}}
\cr}
\eqno(2.1)
$$
where the region surrounding the neighbourhoods $S_\nu$ satisfies the
inequalities $r_\mu(j)\le \vert z-s_\mu(j)\vert \le R_\mu(j)$,
the following condition is imposed on the handles.  For all $j\ge g+1$,
$A_j$ is the homology class represented by the oriented loop
$\phi(\{{\sqrt t_j}e^{i\theta}, {\sqrt t_j}e^{-i\theta}\}\vert 0\le\theta\le 
2\pi)$, $\phi_j: H(t_j)\to Y_j$, then that there exists a 
$\beta > 0$ such that $\sum_{j\ge g+1}~t_j^\beta< \infty$.

The surfaces with nodes should be dense in ${\bar {\mit M}}_\infty$ since the 
addition or subtraction of a handle is a small effect.
If the radii of the bases of the handles in the Euclidean metric of the 
embedding space $r_n$ decrease sufficiently quickly, then the distance in the
moduli space based on the intrinsic metric will tend to zero, and the
original surface would be arbitrarily close to a surface with nodes.  
The condition $\sum_n~r_n^2 <\infty$ implies that $r_n\sim {1\over 
{n^{{1\over 2}+\epsilon}}}$ in the Euclidean metric.  Given that a surface of 
genus $g$ has a hyperbolic metric with curvature -1, so that the total area 
increases linearly, the cross-sectional areas of the handles $\pi r_n^2$ will 
be multiplied by a factor proportional to $n$.  In the intrinsic metric, 
therefore, $r_n^{intr.}\sim {1\over {n^\epsilon}}$, and the distance 
in moduli space between a surface which contains the $n^{th}$ handle and a 
surface for which $r_n^{intr.}\to 0$ will be arbitrarily small.  The property 
of the denseness of surfaces with nodes in infinite-genus moduli space, is 
valid, therefore, since the handles on effectively closed infinite-genus 
surfaces satisfy the above constraint on the cross-sectional areas.  
 
The formulation of string theory in terms of the
analytic geometry of universal moduli space is feasible since the 
process of pinching the surface along an $A_j$-homology cycle and removing 
the nodes allows for a continuous path from the boundary of  ${\mit M}_g$ to 
the moduli space of genus $g-1$ surfaces ${\mit M}_{g-1}$ [3].  The limiting
value of the partition function then exists, so that the factorization
condition, implying that the partition function with nodes equals the
product of the partition functions of the disconnected components, is
well-defined.  Functional derivatives of the partition function, such as
the energy-momentum tensors, also exist.

\vskip 10pt
\noindent
{\bf 2.2. On the Classification of Riemann Surfaces}
\vskip 5pt
There does not exist a Green function for the scalar field with a single delta
function source on a compact surface of finite surface of finite genus
or an $O_G$ surface of infinite genus.
The equation
$$\triangle~G(z,z^\prime)= {1\over {\sqrt {h(z)}}}
                                                \delta(z, z^\prime)
\eqno(2.2)
$$
as 
$\int_{\partial \Sigma}~{\sqrt {h(z)}}\partial_i G(z,z^\prime)dz^i=0$ if the 
boundary $\partial \Sigma$ has zero linear measure. 

Surfaces of $O_{HD}$ type have the property that there exists no non-constant
harmonic function $f(z,{\bar z})$ with finite Dirichlet norm [4]
$$\int_\Sigma~d^2z~{\sqrt{h(z)}}~\vert f^\prime(z)\vert^2
\eqno(2.3)
$$
Over a locally flat coordinate patch, ${\sqrt{h(z)}}$ can be set equal to 1,
and the coordinates in the overlapping neighbourhoods may be chosen so that
this factor equals 1 over the entire domain of the integral.  For a general
choice of coordinates, however, it is necessary to include the ${\sqrt{h(z)}}$.

The class of $O_{AD}$ surfaces are defined by the absence of non-constant 
analytic functions with finite Dirichlet norm.  If no Green function exists, 
then it is not feasible to construct an analytic function by the integral
$$f(z)=\int_\Sigma~d^2z^\prime~G(z,z^\prime)~f(z^\prime)
\eqno(2.4)
$$
Then there is a set of inclusions $O_G\subset O_{HD} \subset O_{AD}$.  

A method for identifying the classification type of the surface 
is the modulus test.  If $\Sigma_n$ is an exhaustive covering of $\Sigma$, 
then the surface belongs to the class $O_{AD}$ if $\prod_{n=1}^\infty~
\mu_n=\infty$ where there is a harmonic function $f$ such that 
$f\vert_{\alpha_{ni}}=0$, $f\vert_{\beta_{ni}}=log~\mu_{ni}$ and  
$\int_{\alpha_{ni}}\ast df=2\pi$, $\alpha_{ni}=E_{ni}\cap \partial 
\Sigma_n$, $\beta_{ni}=E_{ni}\cap \partial \Sigma_{n+1}$, and $\Sigma_{n+1}
-{\bar \Sigma}_n$ consists of finite number of relatively compact regions [5].
The modulus $\mu_n$ is defined to be $min_i~\mu_{ni}$.  
An example is the Schottky covering of a compact genus-$g$ Riemann surface.     
\vskip 10pt
\noindent
{\bf 2.3. Universal Grassmannian}

Consider the action for a free scalar field theory on
a Riemann surface
$$S=\int_\Sigma~d^2z~{\sqrt {h(z)}}\partial X {\bar \partial} X
\eqno(2.5)
$$
The path integral $\int_{\Sigma/D_\Sigma}~D[X] e^{-S}$ can be performed on the
complement of a simply connected compact subset of $\Sigma$ to obtain a
functional of the space of field values on the boundary which is 
homeomorphic to $S^1$.  This functional $\phi_{g,\Sigma}
(X\vert_{bdy(D_\Sigma)})$ can be identified with a state 
$\vert \phi_{g,\Sigma}\rangle$ in the space of Hilbert states on the circle 
[6].

For each Riemann surface $\Sigma$ of genus $g$, the path
integral produces a state $\vert \phi_{g,\Sigma}\rangle$.  Consequently,
there is a direct correspondence between states in the Hilbert space and
compact Riemann surfaces.  Since every compact surface corresponds to 

a particular state through the map
$$\phi: \Sigma \longrightarrow \vert\phi_{g,\Sigma}\rangle
\eqno(2.6)
$$
there is an inclusion
$$\phi: \prod_{g~finite}~{\mit M}_g \rightarrow {\mit H}
\eqno(2.7)
$$
where ${\mit H}$ is the Hilbert space of states of the free field theory
on $S^1$.

This inclusion is not an isomorphism, as there exist 
states in the Hilbert space which are not images of Riemann surfaces under the
map $\phi$.  This can be expected since ${\mit M}_g$ is homeomorphic to a
subset of ${\bf C}^{3g-3}$ and therefore there are linear transformations
do not preserve this set, whereas, the Hilbert space is invariant under these
transformations by definition.
\vskip 10pt
\noindent

{\bf 2.4. Universal Teichm{\"u}ller Space}

A homeomorphism $w:D\to w(D)$ between domains in ${\hat {\bf C}}$ is
quasiconformal if and only if $w$ has locally integrable generalized
derivatives satisfying almost everywhere on $D$ the Beltrami equation
$$w_{\bar z}(z)~=~\mu(z) w_z(z)
\eqno(2.8)
$$
for some measurable complex function $\mu$ on D called the Beltrami 
differential with
$$ess~sup_{z\in D}\vert \mu(z) \vert~=~\vert\vert \mu \vert\vert <1
\eqno(2.9)
$$
By applying the existence and uniqueness theorem to the Beltrami
differential which is $\mu$ on $\Delta$ and is extended to $\Delta^*$
by reflection ${\tilde \mu}\left({1\over {\bar z}}\right)={\bar \mu(z)}
{{z^2}\over {\bar z}}$ for $z\in \Delta$ one obtains the quasiconformal
homeomorphism $w_\mu$ of ${\bf C}$ which is $\mu$-conformal
in $\Delta$, fixes $\pm 1$, and keeps $\Delta$ and $\Delta^*$ invariant.
If the Beltrami differential is G-equivariant, 
$$\mu(\gamma z) {{{\bar\gamma}^\prime(z)}\over {\gamma^\prime(z)}}=\mu(z)
\eqno(2.10)
$$
Let $L^\infty(G)=\{G-compatible~differentials~satisfying~the~Beltrami~ 
equation\}$ [7].  Then the Teichm{\"u}ller space for the Fuchsian group G is
$T(G)=L^\infty(G)/\sim$ where $\mu\sim \nu$ if and only if $w_\mu=w_\nu$
on $\partial\Delta=S^1$.

The universal Teichm{\"u}ller space is obtained when $G=1$, and 
$T(G)\subset T(1)$ for all $G\ne 1$.  Other definitions are
$T(1)=L^\infty(\Delta)_1/\sim$ and $QS(S^1)/SL(2,{\bf R})$.  
A complex analytic model of T(1) is the set of univalent functions in
$\Delta^*$ of the form $F^\mu(z)$ allowing quasiconformal extension to the
whole plane, where $F_\mu(z)=z+{{b_1}\over z}+{{b_2}\over {z^2}}+{{b_3}\over
{z^3}}+...$ corresponds to the conformal equivalence class of the Beltrami
differential $[\mu]$.  The coefficients $b_n$ are coordinates on $T(1)$
and $b_n={\cal O}(n^{-c}),~c=0.509...$ for large $n$ [8].

Any two Beltrami coefficients can be paired so that
$$g(\mu,\nu)=-{{ia}\over {3\pi^2}}\int\int_\Delta \times \int\int_\Delta
{{\mu(z){\bar {\nu(\zeta)}}}\over {(1-z{\bar \zeta})^4}}
d\xi d\zeta \cdot dx dy
\eqno(2.11)
$$
whereas the Weil-Petersson inner product on $T(G)$ is
$$g(\mu,\nu)~=~-{{ia}\over {3\pi^2}}\int\int_{\Delta/G}\times
\int\int_\Delta~{{\mu(z){\bar {\nu (\zeta)}}}\over {(1-z{\bar \zeta})^4}}
d\xi d\zeta
\cdot dx dy
\eqno(2.12)
$$
The integration region $\Delta/G$ is the fundamental domain of the 
Fuchsian group.  In terms of the smooth vector fields on $S^1$,
$v=\sum_m v_m L_m$ and $w=\sum_m w_m L_m$,
$$g(v,w)=-2ia~Re~\sum_{m=2}^\infty~v_m {\bar w}_m~(m^3-m)
\eqno(2.13)
$$
where the infinite series converges if $v$ and $w$ are 
${\bf C}^{{3\over 2}+\varepsilon}$ on $S^1$ [9].

Each map $f\in {\mit T}(G)$ determines a tesselation ${\mit T}_f$ of U which is
invariant under $f\circ G \circ f^{-1}$.  Paramaterizing ${\mit T}_f$ by
a set of glide coefficients $\{k_r^f\}$, $i\in I$, where I is the
index set for the sides of ${\mit T}$ factored by G,  these coordinates can
be used as moduli for Teichm{\"u}ller spaces of finitely and infinitely
generated groups and universal Teichm{\"u}ller space when $G=1$ [10].  

By considering genus-zero surfaces, a presentation can be obtained for
${\mit B}$, a universal modular group.  It can be formulated in terms of
a finite number of relations amongst four operators, the Dehn twist $t$,
the braiding $\pi$, the order 4 rotation $\alpha$ and the order 3 rotation
$\beta$.  Glueing a once-punctured torus $\Sigma_{1,1}$ to each cylinder in 
$\Sigma_{0,\infty}$ defines an infinite-genus surface.  Combining the two
glueing maps $t_s$ with $\pi,~t,~\alpha,~\beta$ one obtains generators of the
universal modular group of an infinite-genus surface $\Sigma_{\infty,\infty}$,
which contains the mapping class groups of surfaces of genus $g$ with $n$
boundary components where $g\ge n$ [11].  Factoring universal Teichm{\"u}ller 
space by this universal modular group gives rise to a space which includes 
the infinite-genus component of universal moduli space.  
\vskip 10pt
\noindent
{\bf 2.5. Extended Schottky Spaces}

The extended Teichm{\"u}ller space ${\bar {\mit T}}_g$ is
$\{(X,\tau)\vert~X~is~a~stable~Riemann~surface~of$
\hfil\break
$genus~g~and~\tau~is~a~standard~set~of~generators~of~\pi_1(X)\}$ [12].  
The quotient of a neighbourhood $U_x$ by the isotropy group $G_x^0$ of
$x$ in the modular group $G_g$ is homeomorphic to an open
submanifold of ${\bar {\mit S}}_g$, the extended Schottky space, so that 
${\bar {\mit T}}_g$ only can be given the structure of a complex ringed space.

The extended Schottky space ${\bar {\mit S}}_g$ which is defined 
to be ${\bar {\mit S}}_g=\{(X,\sigma)\vert~X~stable~Riemann$
$~surface~of~genus~g,\sigma:\pi_1(\Sigma)\to \Gamma~a~homomorphism,~induced~by~
a~Schottky~structure$ $~on~\Sigma\}/\sim$ where 
$(X,\sigma)\sim (X^\prime,~\sigma^\prime)$ if there
is an analytic isomorphism $h:X\to X^\prime$ such that $\sigma=\pi_1(h)\circ 
\sigma^\prime$ up to inner automorphisms of $\Gamma$, the free group 
generated by the $A$-cycles [13]. The extended Schottky space
is a fine moduli space and a complex manifold of dimension 3g-3.
Summation over the genus-$g$ scattering amplitudes could then be
expressed as an integration over the union of extended Schottky spaces
$\prod_{g=0}^\infty~\cup_{k=0}^\infty~{\bar {\mit S}}_g$, which can be mapped
to the universal moduli space through the map $\varphi_g:{\bar {\mit S}}_g\to
{\bar {\mit M}}_g$.

If ${\hat \tau}\in {\mit S}_g$, the image under the map ${\mit S}_g\to {\bf C}^{3g-3}$ is
$$\eqalign{\tau&=\langle K_1,...,K_g, q_2,p_3,q_3,...,p_g,q_g\rangle
\in {\bf C}^{3g-3}
\cr
p_i&=\xi_{1i}~~~~q_i=\xi_{2i}
\cr}
\eqno(2.14)
$$
For dividing nodes, new coordinates [14] must be introduced
$(t_1,...,t_g,\rho_1,...\rho_{2g-3})$, $t_i={1\over {K_i}}$,
$k(j),l(j),m(j),n(j)$ and for $j=1,...,g$, 
$T_k(p_{k(j)})=0$ and $T_k(q_{k(j)})=\infty$
while for $j=g+1,...,2g-3$,~$T_j(p_{k(j)})=0$~$T_j(p_{l(j)})=\infty$,~
$T_j(p_{m(j)})=1$~$\rho_j=T_j(p_{n(j)})$.  New coordinates also would be
introduced at the boundary of super-Schottky space, except that a distinction
must be made between spin nodes of the Ramond and Neveu-Schwarz type. 

The necessity for a change of the coordinates implies that the holomorphic 
measure defined by the Schottky group parameters cannot be extended over 
${\bar {\mit M}}_g$.  The supermoduli space is known to be non-split, and a singular 
transformation [15] is required for the integrand to maintain the form 
$\vert F(y)\vert^2~[det~Im~{\mit T}]^{-5}$ under super-Schottky 
transformations defined by
$${{TZ-Z_{1n}}\over {TZ-Z_{2n}}}=K_n{{Z-Z_{1n}}\over {Z-Z_{2n}}}
\eqno(2.15)
$$
where $K_n$ is the multiplier and $Z_{1m}=(\xi_{1m},\theta_{1m})$, $Z_{2m}=
(\xi_{2m}, \theta_{2m})$ are the super-fixed points.
The integral over the variables $\{K_n,Z_{1m},Z_{2m}\}$ is valid over
supermoduli space except in the neighbourhood of the compactification divisor.
  
\vskip 10pt
\noindent
{\bf 3. Physical Criteria for the Category of Riemann Surfaces}
\vskip 10pt
\noindent
{\bf 3.1. Constraints on the Class of Riemann Surfaces}

The connections between the infinite-genus surfaces and the domain of string
perturbation theory can be analyzed using the concept of flux through the
ideal boundary.  The invariance of the action under the transformation
$X\to X+\epsilon X_n$ gives rise to an infinite nmber of charges $Q_n$ such
that
$$Q_n \vert \phi_{g,\Sigma}\rangle=0~~~~~~~~~~n=1,2,...
\eqno(3.1)
$$
with integrability condition $[Q_m,Q_n]=0$.

At infinite genus, an extra condition on the state is
$$\int_{ideal~bdy}~j_n\vert \phi_{g,\Sigma}\rangle=0 
\eqno(3.2)
$$
Either
\vskip 5pt
\noindent
(i) The ideal boundary has zero harmonic measure.
\hfil\break
(ii) The current $j_n$ vanishes at the boundary.
\hfil\break
(iii) If the ideal boundary is homeomorphic to a circle, the curl of the
current $j_n$ vanishes.
\vskip 5pt
\noindent
Let $X_n(t)=\int^t~[\eta_n-\pi A_{nI}(Im \tau)_{IJ}(\omega-{\bar \omega})_J]$
where $\eta_n$ is a meromorphic differential with only a single pole of
order $n+1$ at $t=0$ and residue $(-n)$ and zero $A$-periods [16] and
$A_{nI}$ is defined by
$$\omega_I=\sum_{n=1}^\infty~A_{nI}t^{n-1}dt~~~~~~~\eta_n=-{1\over {(n-1)!}}
\partial_t \partial_y^n~log~E(t,y)\vert_{y=0}dt
\eqno(3.3)
$$
The current is $j_n=X_n^h{\bar \partial}X+X_n^a\partial X$ where
$$\eqalign{X_n^h~&=~t^{-n}- \sum_{m=1}^\infty 
                                 [2Q_{mn}+\pi A_n (Im~\tau)^{-1}{\bar A}_m] 
                                                       {{{\bar t}^m}\over m}
\cr
X_n^a~&=~\pi \sum_{m=1}^\infty~A_n (Im~\tau)^{-1}{\bar A}_m 
                           {{{\bar t}^m}\over m}
\cr
Q_{nm}~&=~{1\over {2(m-1)!(n-1)!}}\partial_t^m \partial_y^n~{{log~E(t,y)}
\over {(t-y)^{b_n}}}\bigg\vert_{t=y=0}
\cr
b_n&=\left[\sum_{m=1}^\infty~{{(-1)^{m+1}(n)_m}\over {m!}}\right]^{-1}
\cr}
\eqno(3.4)
$$
Since $\partial {\bar \partial}j_n=0$ and the ideal boundary has zero
harmonic measure, the integral condition is immediately satisfied for
$O_G$ surfaces.

If the ideal boundary does not have zero harmonic measure, the values of the 
fields $X_n$, $X$ must tend to zero as $t\to \infty$ for the current 
to vanish.  Then
$X_n^h{\longrightarrow \atop {t\to \infty}} 0$ if and only if
$2Q_{nm}= - A_{nI}(Im~\tau)^{-1}_{IJ}A_{mJ}$, and, indeed, $X_n^h$ 
cannot vanish as $t\to \infty$ unless $A_{nI},~Im~\tau$ 
satisfy this condition.  Similarly, 
$X_n^a{\longrightarrow \atop {t\to \infty}} 0$ if and only if $A_{nI}=0$.  
However, this is not possible because the abelian differentials 
$\omega_I=\sum_{n=1}^\infty~A_{nI}t^{n-1} dt$ do not vanish.  Therefore, 
$X_n^a$ also does not vanish as $t\to \infty$.  Consequently, it is necessary 
to  impose the harmonic condition $\partial {\bar \partial}X=0$ and 
$X(z,{\bar z})\to 0$ as $z$ tends to the ideal boundary.

However, for an $O_{HD}$ surface, there is no non-constant harmonic function
which has finite Dirichlet norm
$$\int_\Sigma~\vert f^\prime(z)\vert^2~dz d{\bar z}~< \infty
\eqno(3.5)
$$
a functional condition which is equivalent to finiteness of the action for
$X(z)$
$$\int_\Sigma~d^2z~\partial X {\bar \partial}X < \infty
\eqno(3.6)
$$
If $X(z)\sim {1\over {\vert z\vert^\varepsilon}}$, $\varepsilon > -1$, and
$\partial X(z)\sim {1\over {\vert z\vert^{1+\varepsilon}}}$
then
$$\eqalign{
lim_{{\mit N}(\partial \Sigma)\to \partial \Sigma}~&\int_{{\mit N}
(\partial \Sigma)}~d^2z~\partial X
{\bar \partial}X~\sim~lim_{z\to \infty}k r_{idl. bdy}
~\int~d\vert z\vert {{\vert z\vert}\over 
{\vert z\vert^{2+2\varepsilon}}}
\cr
&~~~=~lim_{z\to \infty} k r_{idl. bdy}
~\int~{{d\vert z\vert}\over {\vert z \vert^{1+2\varepsilon}}}~=~
-lim_{z\to \infty}~k r_{idl. bdy} 
{{\vert z\vert^{2\varepsilon}}\over {2\varepsilon}}=0
\cr
&~~~k~\le~2\pi
\cr}
\eqno(3.7)
$$
Thus, there are no harmonic functions $X(z)$ satisfying the fall-off condition
$X\sim {1\over {\vert z\vert^\varepsilon}},~\varepsilon> -1$, as they would
have finite Dirichlet norm.  

If $\Sigma\not\in O_{HD}$, the existence of a harmonic function which 
vanishes at the ideal boundary is consistent with the flux condition.   

When the ideal boundary is homeomorphic to a circle,
$$\int_{ideal~bdy}~j_n~dz=~\int_\Sigma~dj_n~d^2z
\eqno(3.8)
$$
and
$$\eqalign{dj_n~&=~(\partial+{\bar \partial})(X_n^h {\bar \partial}X+X_n^a \partial X)
\cr
&=~\partial X_n^h {\bar \partial}X + X_n^h \partial {\bar \partial}X
+\partial X_n^a \partial X + X_n^a \partial^2 X
\cr
&~~~~~~~~~~+{\bar \partial} X_n^h {\bar \partial}X+ X_n^h{\bar \partial}^2X
+{\bar \partial} X_n^a \partial X + X_n^a {\bar \partial}\partial X
\cr
&=~\partial X_n^h {\bar \partial}X + X_n^a \partial^2 X +X_n^h {\bar \partial}^2X
+X_n^h{\bar \partial}^2X
\cr}
\eqno(3.9)
$$
vanishes, in particular, if $\partial X=0$ and ${\bar \partial}X=0$ so that 
$X=constant$ and $dj_n=0$ everywhere on the surface.  The general solution
to the constraint $dj_n=0$ can be obtained by $X$ equal to $\sum_{k,\ell}
~a_{k,\ell} z^k {\bar z}^\ell$ with the coefficients satsifying the recursion
relation
$$\eqalign{&-n(\ell+1)a_{k+n+1,\ell+1}-(\ell+1)\sum_{m=1}^\infty
~[2Q_{mn}+\pi A_n (Im~\tau)^{-1}{\bar A}_m]a_{k+1-m,\ell+1}
\cr
&+(\ell+1)(\ell+2)a_{k+n,\ell+2}-(\ell+1)(\ell+2)\sum_{m=1}^\infty~{1\over m}
[2Q_{mn}+\pi A_n(Im~\tau)^{-1}{\bar A}_m]a_{k-m,\ell+2}
\cr
&+(k+1)(k+2)\pi \sum_{m=1}^\infty {1\over m} A_n (Im~\tau)^{-1} {\bar A}_m
a_{k+2,\ell-m}
\cr
&~~~~~~~~~~~~~~~~~~~~~~~~~~~
~+~(k+1)\pi \sum_{m=1}^\infty A_n (Im~\tau)^{-1} {\bar A}_m a_{k+1,\ell+1-m}
=0
\cr}
\eqno(3.10)
$$

Surfaces which have a boundary at infinity with positive linear measure would 
introduce an additional string state and therefore are excluded from 
the S-matrix expansion, which is defined only for a fixed number of 
asymptotic states.  The domain of string perturbation theory then should 
be restricted to a set of effectively closed surfaces.  Finiteness of the 
size of the interaction region implies that the infinite-genus surfaces must 
be conformally equivalent to spheres of bounded volume with an infinite number 
of handles of diminishing size.  A surface of this kind also can be 
uniformized by an infinitely-generated Schottky group.  When there 
is more than one accumulation point for the handles, the countable sequence 
of generators should be partitioned into subsequences, each corresponding 
to a set of isometric circles with radius decreasing to zero.  

For a planar region with the boundary $\delta$ having capacity 
$c_\delta=e^{-k_\delta}$, the Green function with pole at 
$\infty$ is $G(z,\infty)=log \vert z\vert+k_\beta$.  
Then the Robin constant $r(\delta)=k_\delta$ is shifted to 
$r(\delta^\prime)=r(\delta)+log~a$ under the conformal transformation 
$z=f^{-1}(z^\prime)$ which equals $az^\prime+b+{c\over {z^\prime}}
+{d\over {z^{\prime 2}}}+...$ [5].  Since an orientable Riemann surface can
be conformally mapped to the unit disk, it follows that the property of 
$r(\beta)=\infty$ or equivalently zero capacity of the ideal boundary $\beta$ 
is invariant under conformal transformations.  It may be noted that a 
Stoil{\"o}w ideal boundary point of a noncompact surface, included in the
set $\Sigma^*-\Sigma$, where $\Sigma^*$ is the compactification of $\Sigma$, 
is stable if $c_\gamma=0$ and mapped to a nondegenerate continuum if 
$c_\gamma>0$ [5].  The identification of the class of effectively closed 
surfaces with the category $O_G$ then would be conformally invariant.  
\vskip 10pt
\noindent
{\bf 3.2. String Amplitudes defined by Effectively Closed Surfaces of 
Infinite Genus}

Since string scattering amplitudes includes an integral over the locations
of the vertex operators, they will depend on the Green functions for
fields of various spin on the Riemann surface.  As every closed surface can be 
uniformized by a Schottky group, it can be represented as $\Sigma\sim {\it D}/\Gamma$
where $\Gamma$ is the free group $\langle T_1,..,T_g\rangle$ with $g$ generators
characterized by multipliers $K_n$ and fixed points $\xi_{1n},\xi_{2n}$ through
the relations ${{T_n z-\xi_{1n}}\over {T_n z -\xi_{2n}}}=K_n{{z-\xi_{1n}}\over
{z-\xi_{2n}}}$, $n=1,2,...,g$ and ${\it D}$ is the set of ordinary points points
of $\Gamma$ in the extended complex plane.  When the vertex operators
represent tachyons or dilatons, their correlators are expressed in terms of products of 
exponentials of scalar field Green functions with two 
sources
$$\eqalign{G_{QS}(P,R)~&=~\sum_{\alpha\ne I}~ln \bigg\vert {{z_P-V_\alpha z_R}
\over {z_P-V_\alpha z_S}}{{z_Q-V_\alpha z_S}\over {z_Q-V_\alpha z_R}}
\bigg\vert
\cr
&~~~~~~~~~-{1\over {2\pi}}\sum_{m,n}
~Re\{v_m(z_P)-v_m(z_Q)\}(Im~\tau)^{-1}_{mn}Re\{v_n(z_R)-v_n(z_S)\}
\cr
v_n(z)~&=~\sum_{\alpha\ne I}{}^{(n)}~ln\left({{z-V_\alpha \xi_{1n}}\over
{z-V_\alpha \xi_{2n}}}\right)
\cr
v_n(z)-v_n(T_mz)~&=~2\pi i \tau_{mn}
\cr}
\eqno(3.11)
$$
with $V_\alpha$ being an arbitrary element of the group $\Gamma$ and
$\sum_{\alpha\ne I}{}^{(n)}$ being a sum over elements $V_\alpha$ with the left-most
member of the product of generators not being either $T_n$ or $T_n^{-1}$.
The first term is
$$\sum_{\alpha\ne I}~ln\left[\bigg\vert 1+{{V_\alpha z_S-V_\alpha z_R}
\over {z_P-V_\alpha z_S}}\bigg\vert \bigg\vert 1+{{V_\alpha z_R-V_\alpha z_S}
\over {z_Q-V_\alpha z_R}}\bigg\vert\right]
\eqno(3.12)
$$
and since $\vert V_\alpha z_S -V_\alpha z_R\vert=\vert \gamma_\alpha z_S+
\delta_\alpha\vert^{-1} \vert \gamma_\alpha z_R +\delta_\alpha\vert^{-1}
\vert z_S-z_R\vert$, finiteness of the sum follows from convergence of the
Poincare series $\sum_{\alpha\ne I}\vert \gamma_\alpha\vert^{-2}$.  If 
$r_n\sim {1\over {n^q}}$, $\vert K_n\vert^{1\over 2}=(c_1 n^q+c_2)^{-1}$, 
$\vert \xi_{1n}-\xi_{2n}\vert< c^\prime$, $c< \bigg\vert 
{{\xi_{2n}-{{\alpha_{(l)}}\over {\gamma_{(l)}}}}
\over {\xi_{1n}-\xi_{2n}}}\bigg\vert$, $n=1,2,...$ then
$$\sum_{\alpha\ne I}~\vert \gamma_\alpha\vert^{-2}~<c^2 c^{\prime 2}
\left[{2\over {c_1^2 c^2}}~\sum_{n=1}^\infty~{1\over {n^{2q}}}
~+~\left({2\over {c_1^2 c^2}}~\sum_{n=1}^\infty~{1\over {n^{2q}}}\right)^2
~+~...\right]
\eqno(3.13)
$$
which converges only if $q>{1\over 2}$ [1]. 

The existence of an inverse of the imaginary part of the period matrix, 
$Im~\tau$, follows from the period relations for a compact genus-$g$ surface
$$(\omega, \sigma^\ast)~=~\sum_{k=1}^g~\left[\int_{A_k}\omega
\int_{B_k}{\bar \sigma}~-~\int_{A_k}{\bar \sigma}\int_{B_k} \omega\right]
\eqno(3.14)
$$
where $\omega,~\sigma$ are harmonic differentials and the cycles $A_k,~B_k$
represent a canonical homology basis.  These relations can be generalized to
open manifolds with exhaustion $\{\Omega_n\}$ defined so that there is a
sequence of cycles $A_1,B_1,A_2,B_2,...,A_{p(n)},B_{p(n)},...$ such that
$A_1,B_1,...,A_{p(n)},B_{p(n)}$ form a basis modulo the dividing cycles
of $\Omega_n$
$$(\omega, \sigma^\ast)~=~\sum_{k=1}^\infty~\left[\int_{A_k}\omega
\int_{B_k}{\bar \sigma}~-~\int_{A_k}{\bar \sigma}\int_{B_k} \omega\right]
~+~lim_{n\to \infty}~\int_{\partial \Omega_n}~u{\bar \sigma}
\eqno(3.15)
$$
where $u$ is a function on $\partial \Omega_n$, such that $u(p)=\int_{p_0}^p
\omega$ with $p_0$ being a fixed point on a contour $\alpha$ of
$\partial \Omega_n$ and integration taken to be in the positive sense of 
$\alpha$ [17].  The relations reduce to the usual form for surfaces in the 
class $O_G$ as the ideal boundary has zero harmonic measure.

Holomorphic one-forms and period matrices can be defined generally for
infinite-genus surfaces uniformized by Schottky groups [18] as
$$\eqalign{\omega_j^{(g)}~&=~\sum_\alpha{}^{(j)}~\left({1\over {z-V_\alpha\xi_{1j}}}
-{1\over {z-V_\alpha \xi_{2j}}}\right)dz
\cr
\int_{a_i}~\omega_j^{(g)}~&=~2\pi i \delta_{ij}
\cr
lim_{g\to \infty} \tau_{ij}^{(g)}~&=~{1\over {2\pi i}}lim_{g\to \infty}
                                        \int_{b_i}~\omega_j^{(g)}
\cr}
\eqno(3.16)
$$
The period matrix is
$$\tau_{mn}={1\over {2\pi i}}\left[ln~K_m \delta_{mn}~+~\sum_\alpha{}^{(m,n)}
~ln~\left({{\xi_{1m}-V_\alpha \xi_{1n}}\over {\xi_{1m}-V_\alpha \xi_{2n}}}
{{\xi_{2m}-V_\alpha \xi_{2n}}\over {\xi_{2m}-V_\alpha \xi_{2n}}}\right)
\right]
\eqno(3.17)
$$
where $\sum_\alpha{}^{(m,n)}$ is a sum over all $V_\alpha$ which do not have 
$T_m,T_m^{-1}$ as a left-most member and $T_n, T_n^{-1}$ as a right-most member,
and if $\vert K_n\vert \simeq (c_1 n^q+c_2)^{-2}$, then
$$Im~\tau_{nn}~=~-{1\over {2\pi}}\left[-2q~ln~n~-~2ln~c_1~-~{{2c_2}\over
{c_1 n^q}}\right]~=~{q\over \pi}ln~n~+~{1\over \pi}ln~c_1~+~{{c_2}\over {\pi
c_1 n^q}}
\eqno(3.18)
$$
Given the harmonic functions $v_n(z)=\int^z \omega_n$ [19],
$$\eqalign{Re~v_n(z)~&=~\sum_\alpha{}^{(n)}
~ln\left\vert 1+{{(\xi_{2n}-\xi_{1n})
\gamma_\alpha^{-2}}\over {\left(\xi_{1n}+{{\delta_\alpha}\over {\gamma_\alpha}}
\right)\left(\xi_{2n}+{{\delta_\alpha}\over {\gamma_\alpha}}\right)
(z-V_\alpha \xi_{2n})}}\right\vert
\cr
&\equiv~\sum_\alpha{}^{(n)}~Re~v_{n\alpha}(z)
\cr}
\eqno(3.19)
$$
have the following dependence on $n$, given that $z$ is a bounded distance
$d(z, I_{T_{n_0}})$ from $I_{T_{n_0}}$ for finite $n_0$:
\hfil\break\hfil\break
(i) $d(I_{V_\alpha},I_{T_{n_0}})$, $d(I_{V_\alpha^{-1}}, I_{T_{n_0}})$
are bounded
\hfil\break\hfil\break
\phantom{......}$Re~v_{n\alpha}(z)~=~{\cal O}\left({1\over {n^2}}\right)\vert 
\gamma_\alpha\vert^{-2}$
\hfil\break\hfil\break
(ii) $d(I_{V_\alpha}, I_{T_{n_0}}$, $d(I_{V_\alpha^{-1}}, I_{T_n})$ are
bounded
\hfil\break\hfil\break
\phantom{.......}$Re~v_{n\alpha}(z)~=~{\cal O}\left({1\over {n^3}}\right)\vert 
\gamma_\alpha\vert^{-2}$
\hfil\break\hfil\break
(iii) $d(I_{V_\alpha}, I_{T_n})$, $d(I_{V_\alpha^{-1}}, I_{T_{n_0}})$ are
bounded
\hfil\break\hfil\break
\phantom{.......}$Re~v_{n\alpha}(z)={\cal O}(1)\vert \gamma_\alpha\vert^{-2}$
\hfil\break\hfil\break
(iv) $d(I_{V_\alpha}, I_{T_n})$, $d(I_{V_\alpha^{-1}}, I_{T_n})$ are bounded
\hfil\break\hfil\break
\phantom{.......}$Re~v_{n\alpha}(z)~=~{\cal O}\left({1\over n}\right)
\vert \gamma_\alpha \vert^{-2}$
\hfil\break\hfil\break
when the distances between the isometric circles are bounded.  Given that
$\vert \gamma_n\vert\sim {1\over {n^q}}$, the Poincare series is bounded by
$2{{c^{\prime 2}}\over {c_1^2}}{{\zeta(2q,n)}\over {1-\zeta(2q,n)}}~
{\longrightarrow \atop {n\to \infty}}~{1\over {2q-1}}n^{1-2q}+{\it O}
\left({1\over {n^{2q}}}\right)$ and the sum over the elements in categories 
(iii) and (iv) will decrease at least as ${1\over {n^{2q-1}}}$.  Together 
with the contributions of the elements in categories (i) and (ii), which 
decrease as ${\cal O}\left({1\over {n^2}}\right)$ and 
${\it O}\left({1\over {n^3}}\right)$, the sum over $V_\alpha$ gives 
$Re~v_{n}={\it O}\left({1\over {n^{2q-1}}}\right)$ and  
$Re~\{v_n(z_P)-v_n(z_Q)\}<{{v_{PQ}}\over {n^{2q-1}}}$
for some constant $v_{PQ}$.  Similarly, the dependence of the entries
$(Im \tau)_{mn},~m\ne n$ is $Im~\tau_{mn}~=~{\it O}\left({1\over {\vert m-n
\vert^{2q-1}}}\right)$ so that the eigenvalues of $(Im~\tau)^{-1}$ for large
$n$ are approximately $\lambda_n\simeq {\pi\over q}{1\over {ln~n}}$.  Then
$$\eqalign{{1\over {2\pi}}~\sum_{m,n=1}^\infty&~Re\{v_m(z_P)-v_m(z_Q)\}
(Im~\tau)^{-1}_{mn}~Re\{v_N(z_R)-v_N(z_S)\}
\cr
&< {1\over {2q}}v_{PQ} v_{RS}
\sum_{n=1}^\infty~{1\over {n^{4q-2}~ln~n}}
\cr}
\eqno(3.20)
$$
which is finite for $q>{3\over 4}$.

The integration over the locations of the vertex operators is followed by
a moduli space integral so that the bosonic N-point $g$-loop amplitude [20] is
$$\eqalign{A_g&={{2\pi}\over {(4\pi(8\pi^2)^{13})^g}} 
                            \left({\kappa\over \pi}\right)^{N+2g-2}
\int \prod_{n=1}^g~{{d^2K_n}\over {\vert K_n\vert^4}} \vert 1-K_m\vert^4
\int \prod_{m=1}^g {{d^2\xi_{1m} d^2 \xi_{2m}}\over 
{\vert \xi_{1m}-\xi_{2m}\vert^4}}
\cr
&~~~~~~~~~~~~~~~~~~~~~~~~~~~~~~~~~~~~~~~~~~~~~
\cdot \prod_\alpha{}^\prime~\vert 1-K_\alpha\vert^{-4}
~\prod_\alpha{}^\prime \prod_{p=1}^\infty~\vert 1-K_\alpha^p\vert^{-48}
\cr
&~~~~~~~~~~~~~~~~~~~~~~~~~~~~~~~~~~~~~~~~~~~~
\int~\prod_{s=1}^N~{{d^2z_s}\over {Vol(SL(2,{\bf C}))}}
~\langle V_s(z_s) \rangle
\cr}
\eqno(3.21)
$$
The generalization of the Polyakov measure to the space of surfaces that
can be uniformized by infinitely-generated groups of Schottky type can
be achieved by extending the upper limits of the products over the
multipliers and fixed points to infinity.   

Infinitely-generated Schottky groups only uniformize a specific category
of Riemann surfaces.  If $\Gamma$ is a group of Schottky type with either a 
finite or infinite number of generators with non-overlapping isometric 
circles, and ${\mit D}$ is the set of ordinary points in the extended 
complex plane, then ${{\mit D}\over \Gamma}$ is a Riemann surface in the 
class $O_G$.  The proof is based on a demonstration of the divergence
of the Poincare series for the Fuchsian group uniformizing the surface [1].
Given the existence of a Schottky covering and a universal covering of the
surface by the unit disk, there are conformal equivalences $\Sigma\simeq
{\mit D}/\Gamma\simeq {\tilde{\mit D}}/G$, so that there exists coordinate
neighbourhoods $N_S$ of $z_S\in {\mit D}$ and $N_U$ of $z_U\in {\tilde {\mit 
D}}$ such that there exist homeomorphisms $\Phi_0: N_U\to N_S$ and 
$\Phi_\alpha: {\tilde T}_\alpha N_U\to TN_S$, where
$T\in \Gamma$ is the image of $T_\alpha \in G$ obtained by mapping the 
homology generators $A_1,...,A_g,...$ to the identity.  

It follows that the Poincare series for the Fuchsian group is
$$\sum_{\tilde T\in G}~\vert{\tilde T}^\prime(z_U)\vert
=\sum_{T\in \Gamma}\sum_\alpha~\vert {\tilde T}_\alpha^\prime(z_U)\vert
=\sum_{T\in \Gamma}\sum_\alpha~\vert \Phi_\alpha^{-1}(T\Phi_0(z_U))
\Phi_0^\prime(z_U)\vert \vert T^\prime(z_S)\vert
\eqno(3.22)
$$
 
Let ${\tilde T}_\alpha=A_{i_1}B_{j_1}...A_{i_{N_2}}B_{j_{N_1}}$ where
$$A_{i_1}z_U={{\alpha_{i_1}z_U+\beta_{i_1}}\over 
{\gamma_{i_1}z_U+\delta_{i_1}}}~~~~~~~~~
B_{j_1}z_U={{\alpha_{j_1}z_U+\beta_{j_1}}\over {\gamma_{j_1}z_U+\delta_{j_1}}}
\eqno(3.23)
$$
It follows that
$$\eqalign{\gamma_{A_{i_1}B_{j_1}}&=\gamma_{i_1}
\alpha_{j_1}+\delta_{i_1}\gamma_{j_1}
=\gamma_{i_1}\gamma_{j_1}\left({{\alpha_{j_1}}\over {\gamma_{j_1}}}
+{{\delta_{i_1}}\over {\gamma_{i_1}}}\right)
\cr
\vert \gamma_{A_{i_1}B_{j_1}}\vert^{-2}&=\vert \gamma_{i_1}\vert_{hyp.}^{-2}
\vert \gamma_{j_1}\vert_{hyp.}^{-2}
\left\vert {{\alpha_{j_1}}\over {\gamma_{j_1}}}
+{{\delta_{i_1}}\over {\gamma_{i_1}}}\right\vert_{hyp.}^{-2}
\cr}
\eqno(3.24)
$$
and
$$\eqalign{
\vert \gamma_\alpha\vert^{-2}&=\vert \gamma_{i_1}\vert_{hyp.}^{-2} 
\vert \gamma_{j_1}\vert_{hyp.}^{-2}~...~
\vert \gamma_{i_{N_2}}\vert_{hyp.}^{-2} \vert\gamma_{j_{N_1}}\vert_{hyp.}^{-2}
\cdot \left\vert {{\alpha_{i_{N_2}}}\over {\gamma_{i_{N_2}}}}
+{{\delta_{j_{N_1}}}\over {\gamma_{j_{N_1}}}}\right\vert_{hyp.}^{-2}
\cr
&~~~~\cdot \left\vert {{\alpha_{j_{N_1-1}}}\over {\gamma_{j_{N_1-1}}}}
+{{\delta_{A_{i_{N_2}}B_{j_{N_1}}}}\over {\gamma_{A_{i_{N_2}}B_{j_{N_1}}}}}
\right\vert_{hyp.}^{-2}
~...~\left\vert {{\alpha_{i_1}}\over {\gamma_{i_1}}}
+{{\delta_{B_{j_1}...A_{i_{N_2}}
B_{j_{N_1}}}}\over {\gamma_{B_{j_1}...A_{i_{N_2}}B_{j_{N_1}}}}}
\right\vert_{hyp.}^{-2}
\cr}
\eqno(3.25)
$$
Since $\vert T^\prime(z_S)\vert=\vert \gamma\vert^{-2}\left\vert z_S
+{\delta\over \gamma}\right\vert_{Eucl.}^{-2}$ and
$$\vert \gamma\vert^{-2}=\vert \gamma_{j_1}\vert_{Eucl.}^{-2}...
\vert \gamma_{j_{N_1}}\vert_{Eucl.}^{-2} 
\left\vert {{\alpha_{j_{N_1-1}}}\over {\gamma_{j_{N_1-1}}}}
+{{\delta_{j_{N_1}}}\over {\gamma_{j_{N_1}}}}\right\vert_{Eucl.}^{-2}~...~
\left\vert {{\alpha_{j_1}}\over {\gamma_{j_1}}}+{{\delta_{B_{j_2}...
B_{j_{N_1}}}}\over {\gamma_{B_{j_2}...B_{j_{N_1}}}}}
\right\vert_{Eucl.}^{-2}
\eqno(3.26)
$$
$$\eqalign{\vert {\tilde T}_\alpha^\prime(z_U)\vert &~=~\left\vert z_U+
{{\delta_\alpha}\over {\gamma_\alpha}}\right\vert_{hyp.}^{-2} \left\vert z_S+
{\delta\over \gamma}\right\vert_{Eucl.}^2 \vert \gamma_{i_1}\vert_{hyp.}^{-2}
~...~\vert \gamma_{i_{N_2}}\vert_{hyp.}^{-2}\cdot
{{\vert \gamma_{j_1}\vert_{Eucl.}^2}\over {\vert \gamma_{j_1}\vert_{hyp.}^2}}
~...~{{\vert \gamma_{j_{N_1}}\vert_{Eucl.}^2}\over {\vert \gamma_{j_1}
\vert_{hyp.}^2}}
\cr
&~~~~~~\left\vert{{\alpha_{i_{N_2}}}\over
{\gamma_{i_{N_2}}}}+{{\delta_{j_{N_1}}}\over {\gamma_{j_{N_1}}}}
\right\vert_{hyp.}^{-2}~...~
\left\vert {{\alpha_{i_1}}\over {\gamma_{i_1}}}+{{\delta_{B_{j_1}..A_{i_{N_2}}
B_{j_{N_1}}}}\over {\gamma_{B_{j_1}...A_{i_{N_2}}B_{j_{N_1}}}}}
\right\vert_{hyp.}^{-2}
\cr
&~~~~~~~\cdot \left\vert {{\alpha_{j_{N_1-1}}}\over {\gamma_{j_{N_1-1}}}}
+{{\delta_{j_{N_1}}}\over {\gamma_{j_{N_1}}}}\right\vert_{Eucl.}^2~...~
\left\vert {{\alpha_{j_1}}\over {\gamma_{j_1}}}+{{\delta_{B_{j_2}...
B_{j_{N_1}}}}\over {\gamma_{B_{j_2}...B_{j_{N_1}}}}}
\right\vert_{Eucl.}^2 \vert T^\prime(z_S)\vert
\cr}
\eqno(3.27)
$$
For a given $T\in \Gamma$, represented by a word of length $N_1$, a genus
$g$ can be defined as the number of handles of the smallest compact subset of
the Riemann surface whose fundamental group projects to the minimal subgroup
of $\Gamma$ containing $T$.  The number of irreducible words of length
$N_1+N_2$ in $G$ which project to $T$ is $2g(2g-1)^{N_2-1}{{(N_1+N_2)!}
\over {N_1!N_2!}}$, and since $f(z_U,z_S,\gamma_{i_1},...\gamma_{i_{N_2}},
\gamma_{j_1},\gamma_{j_{N_1}})\prod_{k_1=1}^{N_1+N_2-1}
~d_{k_1}^{-2}~\prod_{k_2=1}^{N_1-1}~d_{k_2}^2~\vert T^\prime(z_S)\vert=
\vert {\tilde T}_\alpha^\prime(z_U)\vert$, where $f$ is a function of the
radii of the arcs and isometric circles for $G$ and $\Gamma$ respectively,
$\{d_{k_1}\}$ are the hyperbolic distances between the centers of the
arcs corresponding to the homology generators $A_{i_{k_1}},k_1=1,...,N_2$,
$B_{j_{N_1+N_2-{k_1}}},k_1=N_2+1,...N_1+N_2-1$ and the center of the arc for
$B_{j_{N_1}}$ in the unit disk and $\{d_{k_2}\}$ are the Euclidean distances
between the centers of the isometric circles,  
$$\eqalign{\sum_{{\tilde T}\in G}~\vert {\tilde T}^\prime(z_U)\vert >&
\sum_{N_1=0}^\infty 
\biggl[\sum_{N_2=0}^\infty~\left({{2g}\over {2g-1}}\right)(2g-1)^{N_2}
f(z_U,z_S,\gamma_{i_1},...\gamma_{i_{N_2}},\gamma_{j_1},\gamma_{j_{N_1}})
\cr
&~~~~~~~~~~~~~~~\prod_{k_1=1}^{N_1+N_2-1}~d_{k_1}^{-2}~
\prod_{k_2=1}^{N_1-1}~d_{k_2}^2~{{(N_1+N_2)!}\over {N_1!N_2!}}\biggr]
~\sum_{T\in \Gamma_{N_1}}\vert T^\prime(z_S)\vert
\cr}
\eqno(3.28)
$$
with $\Gamma_{N_1}$ denoting the elements of $\Gamma$ consisting of the
product of $N_1$ generators or their inverses.
The Euclidean radii of the arcs must tend to zero at least as fast as 
${1\over {n^\alpha}},~\alpha>1$ so that the fundamental domain of the
infinitely generated group G lies within the unit disk.  Since even the 
hyperbolic radii of the some of the arcs may be arbitrarily small, it follows 
that the hyperbolic distances $d_{k_1}$ could tend to zero, leading 
immediately to a divergent product $\prod_{k_1=1}^{N_1+N_2-1}~d_{k_1}^{-2}$.
The product of the distances $\{d_{k_2}\}$ does not affect the divergence
of the sum over $N_2$.  By definition of the parameters $\gamma_{j_1},...,
\gamma_{j_{N_2}}$ and the dilation of the distances in the hyperbolic
metric, the ratios ${{\vert \gamma_{j_1}\vert_{Eucl.}}\over 
{\vert\gamma_{j_1}\vert_{hyp.}}}$,...,
${{\vert \gamma_{j_{N_2}}\vert_{Eucl.}}\over {\vert\gamma_{j_{N_2}}
\vert_{hyp.}}}$ can be set equal to one.  Although the radii 
$\vert \gamma_{i_1}\vert_{hyp.}^{-1},...,
\vert \gamma_{i_{N_2}}\vert_{hyp.}^{-1}$ can decrease to zero, the
distances $d_{k_1}$ simultaneously will tend to zero so that the
product of the radii will be compensated by the product 
$\prod_{k_1=1}^{N_1+N_2-1}~d_{k_1}^{-1}$ if
the final element $B_{j_{N_2}}$ has an arc with an infinitesimally small
radius.  Even when the distances $d_{k_1}$ do not tend to zero,
$(2g-1)^{N_2}\prod_{n=1}^{N_2}~{1\over {n^\alpha}}$ diverges for fixed
$\alpha>1$ as $N_2\to \infty$.  

The infinite factor multiplying the non-vanishing Poincare 
series for the restricted Schottky group $\sum_{T\in \Gamma_{N_1}}
~\vert T^\prime(z_S)\vert$ causes the Poincare series for 
the Fuchsian group to diverge.  Since the uniformizing group 
is a Fuchsian group of the first kind, the surface belongs to 
the class $O_G$.  

For infinite-genus hyperelliptic surfaces, given a diagonal matrix
$T=diag(t_1, t_2,...)$ with $t_j>0,~j\ge 1$ and real numbers
$p>1 $, $C>0$, there is a sequence $\Lambda=\{\lambda_0=0 < \lambda_1 
<\lambda_2 < ... \to \infty\}$ such that the period matrix
$\tau$ of $\Sigma(\lambda)$ is purely imaginary and satisfies 
$\vert(\tau-it)_{ij}\vert \le {C\over {i^p j^p}}t_i^{1\over 2} 
t_j^{1\over 2}$ for all $i,j\le 1$ [21].
Given a sequence of ramification points $\lambda_{2i-1}, \lambda_{2i}\ge Ci^2,
~i\ge 1$, let
$$\eqalign{a_1~&=~2\int_{\lambda_1}^{\lambda_2} {{d\lambda}\over
{\left(\lambda \left({\lambda\over {\lambda_1}}-1\right)\left(1-{\lambda\over
{\lambda_2}}\right)\right)^{1\over 2}}}
\cr             
a_i~&=~{2\over {\lambda_{2i-2}^{1\over 2}}}
\int_{\lambda_{2i-1}}^{\lambda_{2i}} 
{{d\lambda}\over {\left(\left({\lambda\over {\lambda_{2i-2}}}-1\right)
\left({\lambda\over {\lambda_{2i-1}}}-1\right)\left(1-{\lambda\over 
{\lambda_{2i}}}\right)\right)^{1\over 2}}}~~~~~~~~~~i>1
\cr
R(\lambda)~&=~\left(-\lambda\prod_{i=1}^\infty~\left(1-{\lambda\over 
{\lambda_{2i-1}}}\right)\left(1-{\lambda\over {\lambda_{2i}}}\right)
\right)^{1\over 2}
\cr
\varphi_i(\lambda)~&=~a_i^{-1}~\prod_{1\le k\le \infty}~\left(1-{\lambda\over
{\mu_k}}\right)~~~~~~~~\mu_i={\sqrt {\lambda_{2i}\lambda_{2i-1}}}
\cr}
\eqno(3.29)
$$
The holomorphic differentials $\phi_i={{\varphi_i(\lambda)}\over 
{R(\lambda)}}$ with norm $\parallel \phi_i \parallel=\left({i\over 2}
\int_{\bf C}~\biggl\vert{{\varphi_i(\lambda)}\over {R(\lambda)}}\biggr\vert^2
~d\lambda\wedge d{\bar \lambda}\right)^{1\over 2}$ form a basis of smooth, 
closed square-integrable one-forms on $\Sigma$.  If $\omega, \sigma$ are 
linear combinations of $\phi_i$, then
$$\int_\Sigma~\omega \wedge {\bar \sigma}~=~\sum_{i=1}^\infty~\left[
\int_{A_i}\omega~\int_{B_i}{\bar \sigma}~-~\int_{A_i}{\bar \sigma}\int_{B_i}
\omega \right]
\eqno(3.30)
$$
If $A=(A_{ij})_{i,j=1}^\infty,~A_{ij}=\int_{A_i}~\phi_j$, then there exists
a bounded inverse $A^{-1}$ such that $\vert (A^{-1})_{ij}\vert \le \delta_{ij}
+{C\over {i^pj^p}}\rho(i,j)$ and $\psi_j~=~\sum_{k=1}^\infty (A^{-1})_{kj}
\phi_k$ for each $j\in {\bf N}$, the series converges in $L^2$ to a square 
integrable one-form $\psi_j$[21].  The period matrix 
$\tau_{ij}=\int_{B_i}\psi_j$ is symmetric and $Im~\tau$ is positive-definite, 
implying the existence of an inverse.  

The determinant factors in the bosonic string measure on the space of 
hyperelliptic surfaces [22] parameterized by the branch points $\{a_i\}$ are
$$\eqalign{det^\prime {{\bar\partial}_0}&=\prod_{i<j}~(a_i-a_j)^{1\over 4}
\cr
det^\prime {{\bar\partial}_2}&=\prod_{i<j}~(a_i-a_j)^{5\over 4}
\cr}
\eqno(3.31)
$$
so that $\mu_g=(det^\prime {\bar \partial}_0)^{-13}
                     (det^\prime {\bar \partial}_2)=\prod_{i<j} (a_i-a_j)^{-2}$
and
$$Z_g^{hyp.}\sim \int \prod_k d^2a_k \prod_{i<j} \vert a_i-a_j\vert^{-4} 
                    (det~Im~\tau)^{-13}
\eqno(3.32)
$$

The exclusion of surfaces which have fused handles at points other than
the base is apparent in the hyperelliptic representation.  Although 
hyperelliptic surfaces have the property that they possess an involution, so 
that the branch points all lie on the real axis, for example, if the 
$2g+2$ branch points are allowed to have an arbitrary location in the 
extended complex plane, such that the lines joining $a_{2i}$ and $a_{2i+1}$ 
do not intersect for any $i$, then all orientable closed Riemann surfaces 
with a finite number of disjoint handles can be obtained.  

The generalization of the moduli space measure to superstrings introduces
a dependence on the spin structure, which is the choice of square root of
the cotangent bundle and can be viewed as the selection of signs for a
fermion as it traverses around the homology cycles.  The holomorphic part of 
the measure for the Neveu-Schwarz sector [23], consisting of even spin 
structures, is
$${1\over {d{\mit V}_{ABC}}}\prod_{n=1}^g 
{{dK_n}\over {K_n^{3\over 2}}}{{dZ_{1n}dZ_{2n}}\over {K_n^{3\over 2}}} 
\left({{1-K_n}\over {1-(-1)^{B_n}K_n^{1\over 2}}}\right)^2 
(det~Im~{\mit T})^{-5}
\eqno(3.33)
$$
$$
~~~~~\cdot \prod_\alpha{}^\prime \prod_{p=1}^\infty~
\left({{1-(-1)^{N_\alpha^B}K_\alpha^{p-{1\over 2}}}\over 
{1-K_\alpha^p}}\right)^{10} \prod_\alpha{}^\prime \prod_{p=2}^\infty
\left({{1-K_\alpha^p}\over {1-(-1)^{N_\alpha^B}K_\alpha^{p-{1\over 2}}}}
\right)^2
\eqno(3.34)
$$
with
$$d{\mit V}_{ABC}={{dZ_A dZ_B dZ_C}\over 
                       {[(Z_A-Z_B)(Z_C-Z_A)(Z_B-Z_C)]^{1\over 2}}}\cdot 
                       {1\over {d \Theta_{ABC}}}
\eqno(3.35)
$$
$$
\Theta_{ABC}={{\theta_A(Z_B-Z_C)+\theta_B(Z_C-Z_A)+\theta_C(Z_A-Z_B)+\theta_A
                                                    \theta_B\theta_C}
                  \over
                {[(Z_A-Z_B)(Z_C-Z_A)(Z_B-Z_C)]^{1\over 2}}}
\eqno(3.36)
$$
and the super-period matrix is
$${\mit T}_{mn}={1\over {2\pi i}}\left[ln~K_n \delta_{mn}+
\sum_\alpha{}^{(m,n)}~ln\left[{{Z_{1m}-V_\alpha Z_{1n}}\over 
{Z_{1m}-V_\alpha Z_{2n}}}{{Z_{2m}-V_\alpha Z_{2n}}\over 
{Z_{2m}-V_\alpha Z_{1n}}}\right]\right]
\eqno(3.37)
$$
where $Z_{1n},Z_{2n}$ are the superfixed points of the generators of the
super-Schottky group.  The primitive-element products in the measure 
can be derived from the superdeterminants of differential operators
$$\eqalign{sdet^\prime {\bar D}_0~&=~\prod_\alpha{}^\prime
~\prod_{p=1}^\infty~\left({{1-K_\alpha^p}\over 
{1-K_\alpha^{p-{1\over 2}}}}\right)^2
\cr
sdet^\prime {\bar D}_{-1}~&=\prod_\alpha{}^\prime \prod_{p=2}^\infty
\left({{1-K_\alpha^p}\over {1-K_\alpha^{p-{1\over 2}}}}\right)^2
\cr}
\eqno(3.38)
$$
where $D_Z={\partial\over {\partial\theta}}+\theta 
{\partial\over {\partial z}}$.

The superstring amplitude in terms of Fuchsian parameters [24] is
$$\eqalign{Z_g~&=~\int_{sM_g}~d\mu_{SWP}~
[sdet^\prime(-\triangle_0^2)]^{-{5\over 2}}[sdet (-\triangle_2^2)]^{1\over 2}
\cr
&=~\left({1\over {2\pi^4}}\right)^{g-1}
\int_{sM_g}~d\mu_{sWP}~(-1)^{{5\over 2}\triangle n_0^{(0)}}
\left({{Z_0(1)}\over {{\tilde Z}_1({1\over 2})}}\right)^{-5}
{{Z_0(2)}\over {Z_1({3\over 2})}}\left({{Z_1(1)}\over {Z_1(0)}}\right)^2
\cr}
\eqno(3.39)
$$
where the super-Selberg zeta functions are
$$Z_q(s)=\prod_{\{\gamma\}_p}\prod_{k=0}^\infty [1-\chi_\gamma^q 
                    e^{-(s+k)l_\gamma}]~~~~~~~~Re~s>1
\eqno(3.40)
$$
with $q=0,1$, $\chi_\gamma$ denoting the spin structure, $l_\gamma$
being the length of the geodesic corresponding to the primitive element $\gamma$, 
and $\triangle n_0$ equalling the number of even zero-modes minus the number of 
odd zero modes of the Laplace-Dirac operator.
 
By analogy with the bosonic string partition function,  
$$Z_g~=~\int_{M_g}~d\mu_{WP}~Z(2)Z^\prime(1)^{-13}
\eqno(3.41)
$$
where $Z(s)$ is the ordinary Selberg zeta function [25], an expression
for the integrand of the superstring partition function in terms of
branch points has been obtained for the set of hyperelliptic surfaces [26].

A proof of the equivalence of the bosonic string measure in the Fuchsian
and Schottky parameterizations can be given by deriving both expressions
from the path integral
$$Z_g~=~\int~{{[Dh][DX]}\over {Vol(Diff(\Sigma))\times Vol(Weyl(\Sigma))}}
e^{-S(g,X)}
\eqno(3.42)
$$
where $S$ is the Polyakov action.  The evaluation of the determinants of the
Laplacians on genus-g surfaces in the Fuchsian representation gives rise to 
the Selberg zeta functions.  In the Schottky parameterization, the
determinants of the differential operators are obtained on a sphere with
with $g+1$ disks, with the string configurations specified on the boundaries of
the sphere, and then this surface is glued to its Schottky double [27]. 
The equivalence has been demonstrated for manifolds which do not 
have fused handles, which would follow from the hyperbolicity of products of 
hyperbolic generators of the Schottky group $\Gamma$, implying the non-overlap 
of the isometric circles.
\vskip 10pt
\noindent
{\bf 3.3. Dirichlet Boundaries and Non-Perturbative Effects}

Attaching a disk to the Riemann surface, it follows that the non-perturbative
effect of adding a Dirichlet boundary is weighted by a factor of
$exp(-S_{disk})$.  Since $S_{disk}\sim {c\over {\kappa_{str}}}$, the
contribution of the Dirichlet boundary is weighted by
$e^{-{c\over {\kappa_{str}}}}$ [28].

Since gravitational diagrams can be regarded as the square of Yang-Mills
diagrams in $N=8$ supergravity, the coupling constants can be related
as $\kappa_{grav}\sim \kappa_{YM}^2$ [29].  The low-energy limit of the 
non-perturbative contribution to the string amplitude is therefore consistent
with the weighting of non-perturbative effects in non-abelian gauge field
theories, $e^{-{c\over {\kappa_{YM}^2}}}$.

The addition of these Dirichlet boundaries can be viewed as a separate 
contribution to the string path integral.  This formulation of string
theory, then, would be based on a sum over closed finite-genus surface,
effectively closed infinite-genus surfaces and surfaces with Dirichlet 
boundaries inserted.  The first two sums arise within the domain of string
perturbation theory, with only orientable surfaces for closed strings but
non-orientable surfaces also included for open strings.  The last 
sum which represents non-perturbative effects [30] is only well-defined if
the perturbative sum is convergent or can be obtained from analytic
continuation from the domain of convergence. 

\vskip 10pt
\noindent
{\bf 4. Infinite-Genus Surfaces and the String Coupling}
\vskip 10pt
\noindent
{\bf 4.1. The Contribution of Infinite-Genus Surfaces to the Scattering 
Amplitude}

If an infinite-genus surface has a single end, there is an exhaustion by a
sequence of compact surfaces.  In general, the ends may be identified with the
points in a Cantor set defined by the division of the interval.  Defining the
cardinality of the set of ends to be $card~E$, the contribution of 
infinite-genus surfaces to the N-point scattering amplitude
$$\sum_{n_e=1}^{card~E} \sum_{i=1}^{n_e}~lim_{g\to \infty}\kappa_{str}^{2g-2}
\int_{s{\mit M}_{\infty,(i)}}~\sum_{L,L^\prime}~d\mu_{\infty,(i),(L,L^\prime)}
\cdot \prod_{r=1}^N~dt_r d{\bar t}_rV_r(t_r,{\bar t}_r)_{L,L^\prime}
\eqno(4.1)
$$
where $(L,L^\prime)$ denotes the spin structure, $(t_r,{\bar t}_r)$ represent
the local coordinates 
\hfil\break
$(z_r,\theta_r,{\bar z}_r,{\bar \theta}_r)$ on the
super-Riemann surface and $\kappa_{str}$ is the string coupling. 
 
The N-point g-loop graviton Type II superstring amplitude can be determined
in the limit $K_1\sim K_2\sim...\sim K_g\sim 0$ to have the property of
exponentiation in impact space if the coefficient of the integral is
$$C_g^{II}=\left({1\over 4}\right)^g (g_D)^{2g-2}
                    \left({1\over {2\pi}}\right)^{{Dg\over 2}+3g-3}
                     (\alpha^\prime)^{-2+{{g(4-D)}\over 2}}
\eqno(4.2)
$$
where the D-dimensional string coupling is $g_D^2={{2\kappa^2}\over 
{\alpha^\prime}}$, with $\kappa_{D=4}^2=8\pi G_N$ [31].   The same limit 
implies that the coefficient in the bosonic string amplitude is
$$C_g^{bos.}=4^g C_g^{II}
\eqno(4.3)
$$

For open strings, the tension is given by 
$T={1\over {2\pi \alpha^\prime [\ell_P]}}$ where the factor of $[\ell_P]$ is
required for the correct units.  More generally, the tension is
${{m_P c^2}\over {\ell_P^2}}$.  The mass of the closed string 
can be found by assuming uniform density in a circular string with
radius equal to the length of an open string so that $m_P\to 2\pi m_P$.
Because of the oscillatory motion of the string, the effective length of
half of the closed string then would be $2\ell_P$.  The quantity defining
the tension is then transformed to
${{2\pi m_Pc^2}\over {16 \ell_P^2}}={\pi \over 8}T_{open}$ so that
the value of $G_N={1\over {[\ell_P]}}{1\over T}$ is shifted from
$2\pi \alpha^\prime$ to $16\alpha^\prime$.  Then
$g_4^2={{16\pi G_N}\over {\alpha^\prime}}=256\pi,~g_4=16{\sqrt \pi}$, so 
that
$$C_g^{II}=\left({1\over 4}\right)^g g_4^{2g-2}
\left({1\over {2\pi}}\right)^{5g-3}(\alpha^\prime)^{-2}
={1\over {(16{\sqrt \pi})^2 (2\pi)^3 \alpha^{\prime 2}}}
\left({2\over {\pi^4}}\right)^g
\eqno(4.4)
$$
The physical considerations of $\S 3$ imply that the infinite-genus moduli 
space ${\bar {\mit M}}_\infty$ should be identified with the category of $O_G$
surfaces.  Since the boundary of the image of a surface conformally mapped 
to the unit disk is a subset of the circle, the class $O_G$ can be defined 
by constraints on integers $\{p_n\}$, which denote Cantor sets $E(p_1p_2...)$ 
obtained by successively dividing the unit interval into $p_n$ parts and 
deleting the central subinterval.  The generalized Cantor set 
$E(p_1p_2...)=\cap_{n=1}^\infty~E(p_1...p_n)$ has length 
$\prod_{\nu=1}^\infty~\left(1-{1\over {p_\nu}}\right)$ so that the condition 
of zero linear measure is equivalent to $\sum_{\nu=1}^\infty
{1\over {p_\nu}}=\infty$ [5]. 

It has been established that if $r_{n\nu}$ is the Robin constant of 
$E(p_\nu...p_n)$, then $r_n=r_{n1}$ satisfies the inequalities [5]
$$\eqalign{{{log~4}\over {2^n}}&+log\Biggl[\prod_{\nu=1}^n~\left(1-{1\over
{p_\nu}}\right)^{-2^{-\nu}}\Biggr]+\left(1-{1\over {2^n}}\right)log~2
\le r_n
\cr
&~~~~~~~~~~~~~\le {{log~4}\over {2^n}}
 +log\Biggl[\prod_{\nu=1}^n~\left(1-{1\over {p_\nu}}\right)^{-2^{-\nu}}\Biggr]
+\left(1-{1\over {2^n}}\right)log~2+\sum_{\nu=1}^n~{{log~p_\nu}\over {2^\nu}}
\cr}
\eqno(4.5)
$$
As the condition for the Cantor set $E(p_1p_2...)$ to belong to the class
$O_G$ is $lim_{n\to \infty}r_n=\infty$, the integers $\{p_n\}$ would be 
constrained by the relation $\prod_{\nu=1}^\infty~\left(1-{1\over {p_\nu}}
\right)^{2^{-\nu}}=0$ if $\sum_{\nu=1}^\infty~{{log~p_\nu}\over {2^\nu}}$
is finite.  However, since the product equals 
$\left(1-{1\over p}\right)^{\sum_\nu 2^{-\nu}}=1-{1\over p}$ when $p_n=p$
for all $n$, there exists no sequence of positive integers $\{p_n\}$ with
$p_n\le p_{n+1}$ such that $\prod_{\nu=1}^\infty~\left(1-{1\over {p_\nu}}
\right)^{2^{-\nu}}=0$.  However, when $\langle p_\nu\rangle $ is allowed to 
tend to $1$ as $\nu\to \infty$ at a rate of ${1\over {1-k(\nu)}}$, 
where $\langle p_\nu\rangle$ defines an average number of deletions over the 
subintervals of $E(p_1...p_{\nu-1})$, then a generalized Cantor set of
zero capacity would be obtained if $k(\nu)<e^{-\eta^\nu},~\eta> 2$.

An average of the form $\langle p_\nu\rangle={1\over {1-e^{-\eta^\nu}}}$ can 
be obtained if all except one of the subintervals in each of the 
of the partitions at the $(\nu-1)^{th}$ level into $[e^{\eta^\nu}]_{odd},
~n\ge 3$ parts are deleted, including the central section, with 
$[\alpha]_{odd}$ denoting the largest odd integer less than $\alpha$.
This partitioning would yield $lim_{\nu\to \infty}2^\nu$ endpoints.
Different partitionings give rise to endpoint sets which are equivalent
under a conformal transformation.      

Given that the cardinality of the endpoint set is $card~E=2^{\mit N}$, the 
infinite-genus contribution to the scattering amplitude will be multiplied by 
a factor of $lim_{g\to \infty}~2^g$.  The coefficient of the infinite-genus
amplitude then becomes $lim_{g\to \infty}2^g\left({2\over {\pi^4}}\right)^g=
lim_{g\to \infty}\left({4\over {\pi^4}}\right)^g$, which implies that 
the coupling is approximately ${4\over {\pi^4}}\doteq {1\over {24.3523}}$, 
a value which is comparable to the unified gauge coupling 
$\alpha_{GUT}\simeq {{2\pi}\over {153}}={1\over {24.3507}}$
[32][33].

\vskip 10pt
\noindent
{\bf 4.2. The Volume of the Moduli Space of Effectively Closed Infinite-Genus 
Surfaces}

The symplectic volume [34] of the fundamental domain of the modular group 
acting on the Siegel upper half space ${\it H}_g$ is
$$V_g=2^{g^2+1} \pi^{{g(g+1)}\over 2} \prod_{k=1}^g~\left[{{(k-1)!}\over
{(2k)!}}B_{2k}\right]=2 \prod_{k=1}^g (k-1)! \pi^{-k} \zeta(2k)
\eqno(4.6)
$$
The projection of the fundamental domain from ${\it H}_g$ to Teichm{\"u}ller space 
leads to a change in the exponent in the first two factors in the formula 
for $V_g$.  Lowering the dimension of the space from ${1\over 2}g(g+1)$ 
to $3g-3$ implies that $\prod_{k=1}^g (k-1)!\to 
\prod_{k=3}^g 3(k-2)=3^{g-2}(g-2)!$, $\prod_{k=1}^g \pi^{-k}\to 
\pi^{3-3g}$, and $V_g\to 2\cdot 3^{g-2}(g-2)!~\pi^{3-3g}~\prod_{k=1}^g
~\zeta(2k)$.

The differential of the diagonal element of the period matrix is
$$\eqalign{
d\tau_{nn}={1\over {2\pi i}}\biggl[{{dK_n}\over {K_n}}&+\sum_\alpha{}^{(m,n)}
~ln~\left({{\xi_{1m}-V_\alpha\xi_{2n}}\over {\xi_{1m}-V_\alpha \xi_{2n}}}
{{\xi_{2m}-V_\alpha\xi_{2n}}\over {\xi_{2m}-V_\alpha \xi_{1n}}}\right)dK_m
\cr
&+\sum_\alpha{}^{(n,n)}\sum_m{d\over {d\xi_{1m}}}\sum_m~{d\over {d\xi_{1m}}}
ln~\left({{\xi_{1n}-V_\alpha\xi_{1n}}\over {\xi_{1n}-V_\alpha \xi_{2n}}}
{{\xi_{2m}-V_\alpha\xi_{2n}}\over {\xi_{2m}-V_\alpha \xi_{1n}}}\right)d\xi_{1m}
\cr
&+\sum_\alpha{}^{(n,n)}\sum_m {d\over {d\xi_{2m}}}\sum_m~{d\over {d\xi_{1m}}}
ln~\left({{\xi_{1n}-V_\alpha\xi_{1n}}\over {\xi_{1n}-V_\alpha \xi_{2n}}}
{{\xi_{2m}-V_\alpha\xi_{2n}}\over {\xi_{2m}-V_\alpha \xi_{1n}}}\right)d\xi_{2m}
\biggr]
\cr}
\eqno(4.7)
$$
and since
$${d\over {dK_m}} T_m\xi_{1n}={{(\xi_{1n}-\xi_{2m})(\xi_{1m}-\xi_{2m})
(\xi_{1n}-\xi_{1m})}\over {[(\xi_{1n}-\xi_{2m})-K_m(\xi_{1n}-\xi_{1m})]^2}}
\eqno(4.8)
$$
the coefficient of $dK_n$ is of order ${\it O}(1)$.  
The integral
$$\int^{\epsilon^\prime}~{{dx}\over 
{(ln~\left({1\over x}\right))^{g+1}}}=(-1)^{g+1}
{1\over {g!}}li(x)-(-1)^{g+1}x\left[{1\over {g!}}
{1\over {(ln~x)}}+{1\over {g!}}{1\over {(ln~x)^2}}+
...+{1\over g}{1\over {(ln~x)^g}}\right]
\eqno(4.9)
$$
contains a term of maximal magnitude
$${1\over {g(g-1)...(g-k)}}{1\over {\left(ln~\left({1\over {\epsilon^\prime}}
\right)\right)^{g-k}}}
\eqno(4.10)
$$
when 
$${d\over {dk}}\left[{1\over {g(g-1)...(g-k)}}
{1\over {\left(ln~\left({1\over {\epsilon^\prime}}\right)\right)^{g-k}}}\right]
\approx 0
\eqno(4.11)
$$
or equivalently $\psi(g-k+1)-ln\left(ln \left({1\over {\epsilon^\prime}}\right)
\right)=0$ which has solution $k\approx g-ln\left({1\over {\epsilon^\prime}}
\right)$.  The magnitude of the dominant term, when the integrand includes
$[det~Im~\tau]^{-5}$ instead of $[det~Im~\tau]^{g+1}$,
is altered by a factor of $\left({g\over 4}\right)^{g[ln\left({1\over 
{\epsilon^\prime}}\right)]}$ after integration over all $g$ multipliers $K_n$.
    
Since the symplectic form 
$$\wedge_{(i,j)=1}^{{1\over 2}g(g+1)}
{{d\tau_{ij}\wedge d{\bar \tau}_{ij}}\over {[det~(Im~\tau)_{ij}]^{g+1}}}
\eqno(4.12)
$$
yields a volume element containing
$${{d^2K_1}\over {\vert K_1\vert^2}}{{d^2K_2}\over {\vert K_2\vert^2}}
...{{d^2K_g}\over {\vert K_g\vert^2}}{1\over {[det~(Im~\tau)]^{g+1}}}
\eqno(4.13)
$$
finiteness in the $\vert K_n\vert\to 0$ limit follows for both measures.
The change in the integral over $\vert K_n\vert$ from
$\int {{d\vert K_n\vert}\over {\vert K_n\vert \left(ln\left({1\over
{\vert K_n\vert}}\right)\right)^{-(g+1)}}}$
to $\int {{d\vert K_n\vert}\over {\vert K_n\vert \left(ln\left({1\over
{\vert K_n\vert}}\right)\right)^5}}$ provides an extra factor of
$\left({g\over 4}\right)^g (ln~\epsilon^\prime)^{g(g-4)}$.   
The presence of additional terms in the symplectic measure implies that
there is no direct correspondence between the genus-dependence of the
symplectic volume and the superstring integral.

The integral over the superstring measure can be evaluated partially by using 
a parameterization of supermoduli space.  The super-Schottky space can be 
divided into two domains: the first, defined by
limits which are consistent with a genus-independent cut-off for the length 
of closed geodesics on the super-Riemann surface and the second, the remainder 
of the space, representing a neighbourhood of the compactification divisor.

In another set of coordinates on supermoduli space 
$\{K_n, B_m, H_m, {\cal \theta}_{1i},
{\cal \theta}_{2i}\}$, valid over a subset of supermoduli space complementary
to a neighbourhood of the compactification divisor, the holomorphic part of the 
integral over the super-fixed points is replaced by
$$\prod_{m=2}^g~{{dB_m}\over {B_m^{3\over 2}}}~\prod_{m=2}^{g-1} {{dH_m}\over
{H_m^{3\over 2}}}~\prod_{i=2}^{g-1}~
d{\vartheta}_{1i}~\prod_{i=1}^g~d{\vartheta}_{2i}
\eqno(4.14)
$$
Since the genus-g superstring measure has the form $\vert F(y)\vert^2
[sdet \langle \phi \vert \phi \rangle]^{-5}$, where $\{\phi_i\}$ are
super-holomorphic half-differentials [35], modular invariance
implies that $F(y)\sim {1\over y}$ independently of the component of the
boundary of supermoduli space.  The superdeterminant factor
is
$$\eqalign{sdet \langle \phi \vert \phi \rangle &\sim
                               {1\over {ln \vert y\vert}}
~~~~~~~~~~for~an~A-cycle
\cr
sdet \langle \phi \vert \phi \rangle & \sim 1~~~~~~~~~~~~~~~for~a~B-~or~
C-cycle
\cr}
\eqno(4.15)
$$
Under a modular transformation $\sigma_1$ of the surface which maps 
an a $A$-cycle to a $B$-cycle, the dependence is
$$\eqalign{sdet \langle \sigma_1^*\phi \vert \sigma_1^*\phi \rangle 
&\sim 1~~~~~~~~~~~~~~~for~the~A-cycle
\cr
sdet \langle\sigma_1^*\phi \vert \sigma_1^*\phi \rangle &\sim 
{1\over {ln \vert y\vert}}~~~~~~~~~~for~the~B-cycle
\cr}
\eqno(4.16)
$$
and under the transformation $\sigma_2$ from an a $A$-cycle to a $C$-cycle,
$$\eqalign{sdet \langle \sigma_2^*\phi \vert \sigma_2^*\phi \rangle 
&\sim 1~~~~~~~~~~~~~~~for~the~A-cycle
\cr
sdet \langle\sigma_2^*\phi \vert \sigma_2^*\phi\rangle &
\sim {1\over {ln \vert y\vert}}~~~~~~~~~~for~the~C-cycle
\cr}
\eqno(4.17)
$$
Applying the modular transformation from an a A-cycle to a B-cycle, the
holomorphic part of the measure, multiplied by the determinant factor,
changes from
$$\eqalign{
\prod_{n=1}^g~{{dK_n}\over {K_n^{3\over 2}}}& \prod_{m=2}^g {{dB_m}\over
{B_m^{3\over 2}}} \prod_{m=2}^{g-1}{{dH_m}\over {H_m^{3\over 2}}} 
\prod_{i=1}^{g-1} d\vartheta_{1i}
\prod_{i=1}^g d\vartheta_{2i} \left({{1-K_n}\over {1-(-1)^{{\cal B}_n}
K_n^{1\over 2}}}
\right)^2 
\cr
&[det~Im~{\mit T}]^{-5} \prod_\alpha{}^\prime \prod_{p=1}^\infty
\left({{1-(-1)^{N_\alpha^B} K_\alpha^{p-{1\over 2}}}\over
{1-K_\alpha^p}}\right)^{10} \prod_\alpha{}^\prime \prod_{p=2}^\infty
\left({{1-K_\alpha^p}\over {1-(-1)^{N_\alpha^B} K_\alpha^{p-{1\over 2}}}}
\right)^2
\cr}
\eqno(4.18)
$$
to
$$\eqalign{
{{dK_2}\over {K_2^{3\over 2}}}&{{dH_2}\over {H_2^{3\over 2}}}
\prod_{{n=1}\atop {n\ne 2}}^g {{dK_n}\over {K_n^{3\over 2}}}
\prod_{m=2}^g {{dB_m}\over {B_m^{3\over 2}}} 
\prod_{m=3}^{g-1} {{dH_m}\over {H_m^{3\over 2}}}
\prod_{i=1}^{g-1} d\vartheta_{1i}
\prod_{i=1}^g d\vartheta_{2i} \left({{1-K_n(H_2,K_n, \xi_{1n},\xi_{2n})}
\over {1-(-1)^{{\mit B}_n}K_n^{1\over 2}}}\right)^2 
\cr
&\left({{1-H_2}\over {1-(-1)^{{\mit B}_1}H_2^{1\over 2}}}\right)^2
[det~Im~{\mit T}]^{-5}
\prod_\alpha{}^\prime \prod_{p=1}^\infty~\left({{1-(-1)^{N_\alpha^B}
K_\alpha^{p-{1\over 2}}(H_2,K_n,\xi_{1n},\xi_{2n})}\over
{1-K_\alpha^p}}\right)^{10}
\cr
&~~~~~~~~~~\prod_\alpha{}^\prime \prod_{p=2}^\infty~\left({{1-K_\alpha^p
(H_2,K_n,\xi_{1n},\xi_{2n})}\over {1-(-1)^{N_\alpha^B}
K_\alpha^{p-{1\over 2}}}}\right)^2
\cr}
\eqno(4.19)
$$
whereas the modular transformation from an a A-cycle to a C-cycle yields
$$\eqalign{
{{dK_2}\over {K_2^{3\over 2}}}&{{dB_2}\over {B_2^{3\over 2}}}
\prod_{{n=1}\atop {n\ne 2}}^g {{dK_n}\over {K_n^{3\over 2}}}
\prod_{m=3}^g {{dB_m}\over {B_m^{3\over 2}}} 
\prod_{m=2}^{g-1} {{dH_m}\over {H_m^{3\over 2}}}
\prod_{i=1}^{g-1} d\vartheta_{1i}
\prod_{i=1}^g d\vartheta_{2i} \left({{1-K_n(B_2,K_n, \xi_{1n},\xi_{2n})}
\over {1-(-1)^{{\cal B}_n}K_n^{1\over 2}}}\right)^2 
\cr
&\left({{1-B_2}\over {1-(-1)^{{\cal B}_1}B_2^{1\over 2}}}\right)^2
[det~Im~{\mit T}]^{-5}
\prod_\alpha{}^\prime \prod_{p=1}^\infty~\left({{1-(-1)^{N_\alpha^B}
K_\alpha^{p-{1\over 2}}(B_2,K_n,\xi_{1n},\xi_{2n})}\over
{1-K_\alpha^p}}\right)^{10}
\cr
&~~~~~~~~~~\prod_\alpha{}^\prime \prod_{p=2}^\infty~\left({{1-K_\alpha^p
(B_2,K_n,\xi_{1n},\xi_{2n})}\over {1-(-1)^{N_\alpha^B}
K_\alpha^{p-{1\over 2}}}}\right)^2
\cr}
\eqno(4.20)
$$
Finiteness of the integrals of these measures, over the integration domains 
$\sigma_1(F_g)$ and $\sigma_2(F_g)$ respectively, can be verified by evaluating
the $\vert H_m\vert \to 0$ and $\vert B_m\vert \to 0$ limits.

The modular transformation $A_n\to B_n$, $B_n\to -A_n$, which has
the effect of mapping the entry $\tau_{nn}$ of the period matrix to 
$-\tau_{nn}^{-1}$, interchanges the limit $\vert K\vert\to 0$ 
with $\vert K\vert \to 1$.  If $K \to 1$, $\vert K_\alpha \vert \to 1$.  Since
$lim_{K_\alpha\to 1}\left({{1-K_\alpha^p}\over {1-K_\alpha^{p-{1\over 2}}}}
\right)={p\over {p-{1\over 2}}}$, the product of the two 
primitive-element factors for each $\alpha$ in this limit would be
$\prod_{p=1}^\infty~\left({{p-{1\over 2}}\over p}\right)^{10}
\prod_{p=2}^\infty~\left({p\over {p-{1\over 2}}}\right)^2={1\over 2}
\cdot \prod_{p=2}^\infty\left({{p-{1\over 2}}\over p}\right)^8=0$ when 
$N_\alpha^B$ is even.  If $N_\alpha^B$ is odd, the product diverges in the 
limit $K_\alpha\to 1$, but it is well defined if $arg K_\alpha\ne 0$.
The non-zero value of $arg~K_\alpha$ follows from the formula
$$K_{T^2}=K_T^2~\left({{\xi_{1T}+{{\delta_T}\over {\gamma_T}}}\over
{T(\xi_1^{(T^2)})+{{\delta_T}\over {\gamma_T}}}}{{\xi_{1T}+{{\delta_T}\over
{\gamma_T}}}\over {\xi_1^{(T^2)}+{{\delta_T}\over {\delta_T}}}}\right)
\eqno(4.21)
$$
as
$$\eqalign{arg~K_{T^2}&=2~arg~K_T+2~arg\left(\xi_{1T}
+{{\delta_T}\over {\gamma_T}}
\right)-arg\left(T(\xi_1^{(T^2)}+{{\delta_T}\over {\gamma_T}}\right)
-arg \left(\xi_1^{(T^2)}+{{\delta_T}\over {\gamma_T}}\right)
\cr
&=2~arg\left(\xi_{1T}+{{\delta_T}\over {\gamma_T}}\right)
-arg\left(T(\xi_1^{(T^2)})+{{\delta_T}\over {\gamma_T}}\right)
-arg \left(\xi_1^{(T^2)}+{{\delta_T}\over {\gamma_T}}\right)
\cr}
\eqno(4.22)
$$

Divergences in the $H_m$, $B_m$ integrals in the Neveu-Scwharz sector must
be cancelled in a sum over all spin structres. While a modular transformation
maps the entire $\{K_n,H_m,B_m\}$ fundamental domain to another integration
region in this parameter space, the divergences are cancelled amongst each
set of coordinates $\{K_n\}$, $\{H_m\}$ or $\{B_m\}$.  To obtain the
finite remainder, it is sufficient to consider the effect of the modular
transformation on the integrals over each set of coordinates.  
Divergent $H_m$, $B_m$ integrals can be mapped to $K_n$ integrals
in another sector which will yield a finite contribution to the amplitude
upon summation over paired spin structures.  The integrals can be evaluated
given the range of the multiplier in the fundamental domain and its
image under the modular transformation.   

Since ${{\vert \gamma_n\vert^{-2}}\over {\vert \xi{1n}-\xi_{2n}\vert^2}}=
{{\vert K_n\vert}\over {\vert 1-K_n\vert^2}}$, it follows that if 
$\vert \gamma_n\vert^{-2}$ is bounded below, the limits $\vert K_n\vert\to 0$
and $\vert \xi_{1n}-\xi_{2n}\vert \to 0$ cannot be taken simultaneously.
If $\vert \gamma_n\vert^{-2}$ is allowed to vanish, then the limit
$\vert K_n\vert \to 0$ is consistent with a finite value of 
$\vert\xi_{1n}-\xi_{2n}\vert$ and the limit $\vert \xi_{1n}-\xi_{2n}\vert 
\to 0$ is consistent with a finite value of $\vert K_n\vert$.

At genus 3, for instance, there are sequential pairwise cancellations of 
divergences in the $K_n$ integrals in the Neveu-Schwarz sector:
$(-+-+-+),(-+-+--),(-+---+),(-+----) ,(---+-+),(---+--),(-----+),(------)$.
The finiteness of the $K_n$ integrals can be demonstrated for other sectors
obtained by modular transformations of the Neveu-Schwarz and Ramond spin
structures.  The application of the $A_n\to B_n$, $B_n\to -A_n$ 
transformations to the entire set of genus-3 spin structures leads to 
cancellations of divergences in the $H_m$ integrals in the region 
$\sigma(F_3)$.  In this domain, the range of the coordinate $H_m$ is 
determined by the limits by its equivalent role to the $K_m$ variable.  

Denoting the modular transformation which switches $A-$ and $B-$ cycles on
all three handles by $\sigma$, the transformed spin structures may be
listed as
$$\eqalign{\sigma(NS):~~~~(+-+-+-)~~~~~  &
\cr 
                          (+-+---)~~~~~  & 
\cr
                          (+---+-)~~~~~  &
\cr
                          (+-----)~~~~~  &
\cr
                          (--+-+-)~~~~~  &
\cr
                          (--+---)~~~~~  & 
\cr 
                          (----+-)~~~~~  & 
\cr
                          (------)~~~~~  & 
\cr
\sigma(R):~~~~(++++-+)~~~~~   & (++++++)
\cr
(-+-+-+)~~~~~    & (-+-+++)
\cr
(++-+++)~~~~~    & (++-+-+)
\cr
(-+++++)~~~~~    & (-+++-+)
\cr}
$$
$$\eqalign{
\sigma(S_3):~~~~ (+++++-)~~~~~&(++-++-)
\cr
(++++--)~~~~~ & (++-+--)
\cr
(-+-++-)~~~~~ & (-++++-)
\cr
(-+-+--)~~~~~ & (-+++--)
\cr
\sigma(S_4):~~~~ (-++--+)~~~~~ &  (-++-++)
\cr
(-+---+)~~~~~ & (-+--++)
\cr
(+++-++)~~~~~ & (+++-++)
\cr
(++--++)~~~~~ & (++---+)
\cr
\sigma(S_5):~~~~ (+-++-+)~~~~~ & (+-++++)
\cr
(---+-+)~~~~~  &  (--+-++)
\cr
(+--+-+)~~~~~  & (++--+-)
\cr
(--++++)~~~~~  & (--+++-)
\cr
\sigma(S_6):~~~~  (-++-+-)~~~~~ &  (-++---)
\cr
(-++---)~~~~~   & (-+----)
\cr
(-+--+-)~~~~~   & (++--+-)
\cr
(-+----)~~~~~   &  (++----)
\cr}
$$
$$\eqalign{
\sigma(S_7):~~~~ (+-++--)~~~~~ & (---++-)
\cr
(---++-)~~~~~  & (+-++--)
\cr
(---+--)~~~~~  & (--++--)
\cr
(+-+++-)~~~~~  & (--+++-)
\cr
\sigma(S_8):~~~~ (+-+--+)~~~~~ & (+-+-++)
\cr
(+----+)~~~~~  & (+---++)
\cr
(--+--+)~~~~~ & (--+-++)
\cr
(-----+)~~~~~  & (----++)
\cr}
$$
The cancellations in the $H_m$ integrals occur between spin structures in
different sectors, such as $(++++-+)$ in $\sigma(R)$ and $(++++--)$ in 
$\sigma(S_3)$.    
 
Modular transformations of the type $A_n\to B_n,~B_n\to -A_n$ could be used to 
obtain the integrand and domain for the odd spin structures given an
expression for the Ramond measure.  An odd spin structure can 
be obtained from an even spin structure with a genus-one 
component $(--)$ by changing the periodicity of the fermion 
from $\psi\to -\psi$ to $\psi\to \psi$ upon traversal of either the 
A- or B-cycle, so that the spin structure for this component is then $(++)$.
This transformation maps the spin structures of the Neveu-Schwarz sector
$(-\sigma_{h_1}~...~-\sigma_{h_g})$, $\sigma_{h_i}=\pm 1,i=1,...,g$  
to the spin structures of the Ramond sector $(+(-\sigma_{h_1})~...~ 
+(-\sigma_{h_g}))$.  This can be viewed as a coordinate change 
on the super-Riemann surface $z\to z,~\theta\to -\theta$.  
Under $SPL(2;{\bf C})$ transformation [24]
$$\eqalign{z^\prime&={{az+b}\over {cz+d}}
                           +\theta {{\alpha z+\beta}\over {(az+b)^2}} 
\cr
\theta^\prime&={{\alpha+\beta z}\over {cz+d}}+{{\chi_T~\theta}\over {cz+d}}
\cr
ad-bc&=1+\alpha\beta
\cr
sdet~T&=\chi_T=\pm 1
\cr}
\eqno(4.23)
$$
$(z,\theta)\to (z,-\theta)$ if $a=1,b=0,c=0,d=1,\alpha=0,\beta=0,\chi_T=-1$.
Since this transformation has superdeterminant -1, its product with a
super-Schottky group transformation cannot be expressed in the form
$${{{\tilde T}Z-{\tilde Z}_{1n}}\over {{\tilde T}Z-Z_{2n}}}={\tilde K}_n
{{Z-{\tilde Z}_{1n}}\over {Z-{\tilde Z}_{2n}}}
\eqno(4.24)
$$
The direct substitution of the variables $(K_n,Z_{1n},Z_{2n})\to 
({\tilde K}_n,{\tilde Z}_{1n},{\tilde Z}_{2n})$ cannot be used to
define the measure for the spin structure with the opposite signs for
periodicity properties of the fermion.

It is therefore preferable to approximate the superstring path integral
either by summing over distinct triangulations of genus-$g$ surfaces or
dominant contributions determined by extremizing the action.
Triangulations of a Riemann surface lead to the following estimate [36] 
$$Vol[{\mit M}(T_a)]\rightarrow_{{\mit V}vol(\sigma^2)\ll 1}
                                               {\tilde A}_h({\it V})
                                                       \cdot N_2^{{5\over 4}
                                                      \chi(\Sigma)-{5\over 2}}
\eqno(4.25)
$$
where ${\mit V}\doteq vol\left[{{Hom(\pi_1(\Sigma,G))}\over G}\right]$,
$\sigma^2$ denotes the 2-simplex,
${\tilde A}_h({\mit V})$ is a constant and
$N_2$ is the number of two-dimensional simplices, which satisfies the
constraint
$$N_0-N_1+N_2=\chi(T)
\eqno(4.26)
$$
A typical triangulation of a genus-$g$ surface is given by replacing the handle
by a tetrahedron with the following number of simplices
$$\eqalign{N_0&=g+2
\cr
N_1&=12g
\cr
N_2&=9g
\cr
N_0-N_1+N_2&=2-2g
\cr}
\eqno(4.27)
$$
These values gives rise to a genus-dependence ${\tilde A}_h({\it V})
(9g)^{{1\over 2}-2g}$, which represents a decreasing portion of the
entire moduli space as $g\to \infty$.  With increasing genus, the condition
${\it V}vol(\sigma^2)\ll 1$ limits the moduli space to triangulations
by simplices which can be included in a sequence of embeddings
$...\subset S^{n-4}\subset S^{n-2} \subset S^n \subset ...$ with
$vol(\sigma^2)\sim vol(S^2)\le vol(S^n){{\longrightarrow}\atop {g\to \infty}}
0$.    

The weighting factor in the partition function for two-dimensional Lorentzian 
gravity, $e^{-S_{grav.}}$, has an exponential dependence on the number of
handles in the manifold.  A similar result would be expected for the 
bosonic string action which is bounded below by the area of the worldsheet 
and equals the minimum value when the intrinsic metric is conformal to 
$\partial_\alpha X^\mu \partial_\beta X_\mu$.
Since the femionic string action [37]

$$\eqalign{
S(E,\Phi)&={1\over {2\pi \alpha^\prime}}
\int d^2z d^2\theta e E_\alpha^M \partial_M \Phi(Z) E^{\alpha N}
\partial_N \Phi(Z)
\cr
S(h,X,\chi,\psi)&={1\over {2\pi\alpha^\prime}}\int_\Sigma~d^2z {\sqrt h}\biggl[
{1\over 2}h^{\alpha\beta}
\partial_\alpha X^\mu \partial_\beta X_\mu-{1\over 2}i{\bar \psi}^\mu 
\rho^\alpha \partial_\alpha \psi_\mu
\cr
&~~~~~~~~~~~~~~~~~~~~+{\bar \chi}_a \rho^\alpha
\rho^a \psi^\mu \partial_\alpha X_\mu + {1\over 4}{\bar \psi}^\mu \psi_\mu
{\bar \chi}_a \rho^a\rho^b \chi_b\biggr]
\cr}
\eqno(4.28)
$$
has equation of motion $D_\alpha{\bar D}^\alpha \Phi(Z)=0$, $D_\alpha=
E_\alpha^M\partial_M$, for the coordinate superfield $\Phi(Z)$,
the extremal value is given by the embedded area in superspace 
determined by harmonic maps $z\to \Phi(Z)$.

It is known that the conformal factor in the metric $ds^2=e^\phi dz d{\bar z}$
is approximately ${1\over {r^2(ln~r)^2}}$ near a parabolic puncture [38], with
$r$ being the distance parameter.  The equation of motion for 
$h^{\alpha\beta}$ implies that $h_{\alpha\beta}=\partial_\alpha X^\mu 
\partial_\beta X_\mu$ so that $\partial_z X^\mu \partial_{\bar z}
X_\mu={1\over {2r^2(ln~r)^2}}$ defines a configuration which 
gives a dominant contribution to the string path integral.  A
solution to this equation is $X^\mu=X_0^\mu+{1\over 2}~ln(ln~r){\bf e}^\mu$.   
The spinor field then will take the form 
$\psi_0^\mu-i\rho^\alpha \partial_\alpha X^\mu\epsilon=
\psi_0^\mu-i{1\over {4r^2ln~r}}(\rho^z {\bar z}+\rho^{\bar z}z){\bf e}^\mu\epsilon$.
A gauge can be chosen so that $e^\alpha_a=\delta^\alpha_a,~\chi_\alpha=0$,
implying that only the first two terms of $S(h,X,\chi,\psi)$ would be
non-vanishing.  Substituting the functional dependence of $X^\mu$ and 
$\psi^\mu$ into the action 
$$\eqalign{\sum_i~&\int_{{\mit N}(Q_i)}
d^2z~{\sqrt h}\left[{1\over 2}h^{\alpha\beta}
\partial_\alpha X^\mu \partial_\beta X_\mu-{1\over 2}i{\bar \psi}^\mu
\rho^\alpha\partial_\alpha\psi_\mu\right]
\cr
&=\sum_i~\int_{{\mit N}(Q_i)}~d^2z_i\biggl[{1\over 2}
{1\over {r_i^2(ln~r_i)^2}}
-{1\over 2}i\left[{\bar \psi}_0^\mu+{i\over {4r_i^2~ln~r_i}}{\bar \epsilon}{\bf
e}^\mu
(\rho^{z_i}z_i+\rho^{{\bar z}_i}{\bar z}_i)\right]
\cr}
$$
$$
~~~~~~~~~~~~~~~~~~~~~~~~~~~~~~~~~~~~~~~~~~~~~~~~~~~~~~~~~~~~~~~~~
\rho^\alpha \partial_\alpha
\left[\psi_0^\mu-{i\over {4r_i^2ln~r_i}}(\rho^{{\bar
z}_i}z_i+\rho^{z_i}{\bar
z}_i)
{\bf
e}_\mu
\epsilon \right]\biggr]
\eqno(4.29)
$$
where $Q_i$ denotes the location of the puncture on the surface.
Setting
the
average
value
of
the
fermion
in
the
background
field
$\langle\psi_0^\mu\rangle$
equal
to
zero,
the oscillatory factor can be discarded in an evaluation of the magnitude of the
path integral leaving
$$\eqalign{\int~{\it D}(h,X,\psi)~e^{-S(h,X,\chi,\psi)}
\sim &\sum_{worldsheets}~e^{-{1\over {2\alpha^\prime}} \sum_N\sum_i
~\int_{{\mit N}(Q_i)}~{{dr_i}\over {r_i~(ln~r_i)^2}}}
\cr
&~~~~~~~=\sum_{worldsheets}~e^{-{1\over {2\alpha^\prime}} \sum_N\sum_i~ 
-{1\over {(ln~r)}}\vert_{\partial({\mit N}(Q_i))}}
\cr}
\eqno(4.30)
$$
where the sums are defined by the decomposition of the Riemann surface into
$2g-2$ three-punctured spheres [39].  Given that the  maximum distance between
three punctures on a sphere is ${{2\pi}\over 3}R$, where $R$ is the radius of
the sphere, the contribution to the string path integral is
$e^{{3\over {\alpha^\prime}}(g-1)\cdot \left(log~{\pi\over N}R
\right)^{-1}}$ multiplied by a combinatorial factor equal to the number 
of different ways of linking the 2$g$-2 spheres with each element of the set 
$I_1=\{S_2,...,S_{g-1}\}$ connected to three spheres, the spheres $S_3,...,S_{g-2}$ 
having two adjacent connections in $I_1$, and every element of the set 
$I_2=\{S_1,S_g,...,S_{2g-2}\}$ linked only to one of the spheres in the set $I_1$.  
The combinatorial factor is then $\left({4\atop 2}\right)\left({{2g-4}\atop {g-2}}\right)
\longrightarrow_{g\to \infty} 6\cdot 2^{2g-4}$.  

Since $\ell_P\simeq 4R$, $R\sim (2\pi)^{1\over 3}$ and the contribution of the
infinite-genus surfaces to the vacuum amplitude becomes
$$\eqalign{lim_{g\to \infty}&{3\over 2}\kappa_{str}^{2g-2}\cdot 8^g \cdot
e^{6(g-1)\left(ln({{\pi{(2\pi)^{1\over 3}}}\over N}\right)^{-1}}
\cr
&=lim_{g\to \infty}{3\over {2\kappa_{str}^2}}
e^{-{6\over {ln~\left({{\pi{(2\pi)^{1\over 3}}}\over N}\right)}}}
\left(8\kappa_{str}^2 \cdot e^{6\over {ln~\pi{(2\pi)^{1\over 3}}-ln~N}}
\right)^g
\cr}
\eqno(4.31)
$$
Setting the base of the exponential equal to 1, it follows that
$$\kappa_{str}^2={1\over 8}e^{6\over {ln~N-1.84527}}
\eqno(4.32)
$$
Suppose that $\kappa_{str}\simeq {1\over {24.3507}}$, a condition derived from 
the equality of $\kappa_{str}$ and the unified gauge coupling constant $\alpha_{GUT}$.  
Then $N=e^{\left(1.84527+{6\over {ln~0.0134917}}\right)}\simeq 1.57107$.
The integration range for the spherical radius of the neighbourhood of the 
puncture with metric ${1\over {r^2(ln~r)^2}}dz d{\bar z}$ would extend
to ${\pi\over {1.57107}}R$ or approximately ${{2\pi}\over 3}R$.

\vskip 10pt
\centerline{\bf Acknowledgements}
\noindent
This work has been completed with financial support of the 
Sonderforschungsbereich 288 `Differential Geometry and Quantum Physics'.
The interest of Prof. R. Schrader is gratefully acknowledged.

\vfill\eject
\centerline{\bf References}
\item{[1]} S. Davis, Class. Quantum Grav. ${\underline 6}$(1989)1791
\item{[2]} D. Friedan and S. Shenker, Nucl. Phys. ${\underline{B281}}$ (1987)
                                                                  509
\item{[3]} J. Feldman, H. Knoerrer and E. Trubowitz, 
${\underline{Infinite-Genus~Riemann~Surfaces}}$, Canadian Mathematical Society,
Volume 3, Invited Talks, Ottawa, 1996
\item{[4]} L. Ahlfors and Sario, ${\underline{Riemann~Surfaces}}$,
Princeton Mathematical Series 26 (Princeton: Princeton University Press, 1960)
\item{[5]} L. Sario and M. Nakai, ${\underline{Classification~Theory~of 
~Riemann~Surfaces}}$ (Berlin: 
\hfil\break
Springer-Verlag, 1970)
\item{[6]} L. Alvarez-Gaume and C. Reina, Phys. Lett. ${\underline{190B}}$
(1987)55
\item{[7]} L. Bers, Bull. London Math. Soc. ${\underline{4}}$(1972)257
\item{[8]} O. Pekonen, J. Geom. Phys. ${\underline{15}}$(1995)227
\item{[9]} S. Nag and A. Verjovsky, Commun. Math. Phys. ${\underline{130}}$
(1990)123
\item{[10]} P. Tukia, Ann. Acad. Sci. Fenn. Ser. AI ${\underline{3}}$(1977)343
\item{[11]} L. Funar and C. Kapoudjian, `On Universal Mapping Class Groups',
math.GT/0210007
\item{[12]} F. Herrlich, Math. Z. ${\underline{203}}$(1990)279
\item{[13]} L. Gerritzen and F. Herrlich, J. Reine Angew. Math. 
${\underline{389}}$(1988)190
\item{[14]} H. Sato, Nagoya Math. J. ${\underline{88}}$(1982)79
\item{[15]} G. Danilov, Phys. Atom. Nucl. ${\underline{60}}$ (1997) 1358
\item{[16]} C. Vafa, Phys. Lett. ${\underline{190B}}$(1987)47
\item{[17]} R. D. M. Accola, Trans. Amer. Math. Soc. ${\underline{96}}$
(1960)143
\item{[18]} T. Ichikawa, J. Reine Angew. Math. ${\underline{486}}$(1997)45
\item{[19]} S. Davis, Mod. Phys. Lett. ${\underline{A9}}$(1994)1299
\item{[20]} H. Ooguri and N. Sakai, Nucl. Phys. ${\underline{B312}}$(1989)435
\item{[21]} W. M{\"u}ller, M. Schmidt and R. Schrader, Duke Math. J. 
${\underline{91}}$(2)(1998)315
\item{[22]} H.-S. La, Ann. Phys. ${\underline{205}}$(1991)458
\item{[23]} A. Bellini, G. Cristofano, M. Fabbrichesi and K. Roland,
Nucl. Phys. ${\underline{B356}}$(1991)69
\item{[24]} C. Grosche. Commun. Math. Phys. ${\underline{133}}$ (1990) 433
\item{[25]} E. D'Hoker and D. H. Phong, Rev. Mod. Phys. ${\underline{60}}$
(1988) 917 
\item{[26]} D. Lebedev and A. Morozov, Nucl. Phys. ${\underline{B302}}$
(1988) 163
\item{[27]} M. A. Martin-Delgado and J. Ramirez Mittelbrun, Int. J. Mod. Phys.
${\underline{A6}}$ (1991) 1719
\item{[28]} J. Polchinski, Phys. Rev. ${\underline{D50}}$(1994)6041
\item{[29]} Z. Bern, L. J. Dixon, D. C. Dunbar, M. Perelstein and 
J. S. Rosowsky, Nucl. Phys. ${\underline{B530}}$(1998)401
\item{[30]} M. Green and P. Wai, Nucl. Phys. ${\underline{B431}}$(1994)131
\item{[31]} A. Pasquinicci and K. Roland, Nucl. Phys. ${\underline{B457}}$
(1995)27
\item{[32]} B. de Carlos, J. Casas and C. Munoz, Nucl. Phys. 
${\underline{B399}}$ (1993)623
\item{[33]} M. Shifman, Int. J. Mod. Phys. ${\underline{A11}}$ (1996) 5761
\item{[34]} C. L. Siegel, ${\underline{Symplectic~Geometry}}$ 
(New York: Academic Press, 1964)
\item{[35]} K. Nishimura, Prog. Theor. Phys. ${\underline{84}}$(1990)360
\item{[36]} J. Ambjorn, M. Carfora, A. Marzuoli, The Geometry of
Dynamical Triangulations, hep-th/9612069
\item{[37]} M. Kaku, Strings, Conformal Fields and M-Theory (New York:
Springer, 2000)
\item{[38]} L.Takhtajan, Mod.Phys.Lett. ${\underline{A8}}$ (1993)3529
\item{[39]} N. Berkovits, Nucl. Phys. ${\underline{B408}}$ (1993)43

\end